\documentclass[preprint,11pt]{elsarticle}

\usepackage[margin=1.1in]{geometry}
\usepackage{amssymb}
\usepackage{graphicx}
\usepackage{physics}
\usepackage{amsmath}
\usepackage{derivative}
\usepackage[version=4]{mhchem}
\usepackage{siunitx}
\usepackage{longtable,tabularx}
\usepackage{floatrow}
\usepackage{url}
\usepackage{hyperref}
\usepackage{comment}
\usepackage{todonotes}
\usepackage[table]{xcolor}

\definecolor{smallred}{RGB}{255,220,220}
\definecolor{largegreen}{RGB}{220,245,220}

\begin{document}

\begin{frontmatter}



\title{Mechanisms of lift generation and drag invariance by asymmetric surface roughness on a sphere}

\author[inst1]{Putu Brahmanda Sudarsana}
\author[inst2]{Jagmohan Singh}
\author[inst1,inst3]{Anchal Sareen}

\affiliation[inst1]{organization={Department of Naval Architecture and Marine Engineering, University of Michigan},
            city={Ann Arbor},
            postcode={48109}, 
            state={MI},
            country={USA}}

\affiliation[inst2]{organization={Department of Aerospace Engineering, University of Michigan},
            city={Ann Arbor},
            postcode={48109}, 
            state={MI},
            country={USA}}

\affiliation[inst3]{organization={Department of Mechanical Engineering, University of Michigan},
            city={Ann Arbor},
            postcode={48109}, 
            state={MI},
            country={USA}}

\begin{abstract}

The mechanisms governing lift or transverse force generation on a sphere with asymmetric dimpled roughness are investigated using wall-resolved large eddy simulation at Reynolds numbers, $Re = U_\infty d/\nu$ = $100{,}000$, where $U_\infty$ is the freestream velocity, $d$ is the sphere diameter, and $\nu$ is the kinematic viscosity. The dimple depth ratios of $k/d = 0.004$, $0.006$, and $0.008$ are investigated, where $k$ is the dimple depth. Previous experiments by Sudarsana et al. \cite{sudarsana2024} showed that asymmetric surface roughness can generate lift comparable to the peak Magnus force on a rotating sphere, while leaving the mean drag nearly unchanged. However, the mechanisms underlying the near-invariant drag and the inherently three-dimensional separation dynamics could not be resolved experimentally. The present simulations reproduce the experimentally observed force characteristics and provide the detailed characterization of the flow physics governing lift generation and drag invariance over an asymmetrically dimpled sphere. A spatial decomposition of the surface pressure reveals that the asymmetric dimple-induced perturbation redistributes the streamwise pressure contribution between the upstream and downstream hemispheres with a small change in the net drag, while simultaneously generating a finite transverse pressure difference that produces lift. The invariance of streamwise-projected pressure distribution across different $k/d$ is further elaborated by decomposing the surface pressure field into axisymmetric and non-axisymmetric terms using Fourier decomposition. Analysis of the three-dimensional separation topology reveals two coexisting transition pathways. On the dimpled hemisphere, the boundary-layer undergoes transition upstream of separation, delaying flow separation non-uniformly to $\phi_s\sim105^\circ$--$125^\circ$, where $\phi_s$ denotes the azimuthal separation angle. In contrast, the boundary layer separates in a laminar state at $\phi_s\sim80^\circ$ on the smooth hemisphere. The resulting pressure asymmetry drives a sidewash that rolls up into a counter-rotating streamwise vortex pair, analogous to the tip vortices of a finite wing. This vortex pair amplifies the wake deflection well beyond that is expected from the change in the separation angle alone. The results demonstrate that lift generation on an asymmetrically dimpled sphere arises from a complex and previously unresolved sequence of coupled flow phenomena, whereby localized asymmetric roughness fundamentally alters the three-dimensional separation topology and wake organization. Lift is generated through the coupled interaction of asymmetric boundary-layer transition, non-uniform separation delay, and the formation of a counter-rotating streamwise vortex pair, collectively linking a localized surface perturbation to a global reorganization of the wake while leaving the mean drag nearly unchanged. The present findings provide a mechanistic framework for understanding and exploiting surface-induced lift generation in three-dimensional separated flows.

\end{abstract}


\end{frontmatter}

\section{Introduction}\label{sec1}

Flow over a sphere represents a canonical bluff-body problem that exhibits a rich variety of flow phenomena across different Reynolds numbers, including boundary-layer transition, flow separation, wake instabilities, and abrupt drag reduction. Owing to its geometric simplicity and rich flow physics, the sphere has long served as a fundamental model system for understanding bluff-body aerodynamics. Foundational experiments by Achenbach \cite{Achenbach1972} on flow around a sphere categorize the flow behavior into four distinct regimes: subcritical, critical, supercritical, and transcritical. This classification is based on the variation of the drag coefficient, $C_D$, across different Reynolds numbers $Re$. At subcritical $Re$, the boundary layer separates in a laminar state and remains separated, with $C_D$ nearly constant at $\sim0.5$. As the flow enters the critical regime ($Re\approx3.7\times10^5$), the detached layer becomes turbulent and reattaches to the surface. This generates a laminar separation bubble (LSB) that significantly delays the global flow separation \cite{Achenbach1972, Deshpande2018, Parekh2024}. The delayed flow separation significantly reduces the wake size leading to a sudden drop in drag by up to 50\%, known as the drag crisis. In the supercritical regime, the boundary layer transitions to turbulence before separating, without forming an LSB, while at transcritical $Re$, the transition point in the boundary layer shifts closer to the stagnation point, accompanied by an increasing $C_D$ \cite{Achenbach1972}. An addition of surface roughness has been shown to reduce the critical $Re$ at which the drag crisis occurs, with a higher roughness parameter resulting in a lower critical $Re$ \cite{Achenbach1974a}. Dimples also behave similarly by transitioning the detached shear layer and delaying the global flow separation on spheres \cite{Choi2006, Choi2008, Beratlis2019}. However, the dimpled sphere maintains its drag-crisis value in the supercritical regime as $Re$ increases \cite{Choi2006, Choi2008}, whereas protruding roughness significantly increases the drag above the drag-crisis value \cite{Achenbach1974a, Beratlis2019}. For this reason, dimples are widely regarded as an effective surface roughness for reducing drag on spheres and have therefore been extensively investigated as an effective flow-control strategy for bluff-body drag reduction \cite{Choi2006, Chae2026, Terwagne2014, vilumbrales2025}. While symmetric roughness distributions primarily modify the drag characteristics of the sphere, asymmetric surface perturbations can fundamentally alter the three-dimensional separation topology and generate substantial transverse forces.


 
Transverse force generation on a sphere has long been associated with asymmetric boundary-layer states. In the classical Magnus effect, sphere rotation alters the relative boundary layer development on the two sides of the body, shifting the separation location asymmetrically and producing a transverse force \cite{Muto2012, Krishnan2025, Milner2025}. The underlying mechanism is strongly tied to asymmetry in boundary layer transition, separation, and wake pressure recovery \cite{mehta1985, Kim2014, Muto2012, Milner2025}. Similar physics also governs the aerodynamics of sports balls, particularly the cricket ball, where the angled seam promotes earlier transition on one side of the ball while the opposite side remains relatively smooth. The resulting asymmetric boundary-layer development produces an imbalance in the surface pressure distribution and a corresponding transverse force, commonly referred to as swing \cite{mehta1985, Deshpande2018, Parekh2024}. Deshpande et al. \cite{Deshpande2018} further showed that cricket-ball swing depends strongly on $Re$ and can be classified into three regimes: no swing, conventional swing, and reverse swing. In the no-swing regime, the seam does not produce sufficient perturbation to alter the boundary layer state on the seam side. Beyond a critical $Re$, the seam-side boundary layer becomes sufficiently perturbed by the seam to generate a transverse force, leading to conventional swing. With further increase in $Re$, the boundary layer on the smooth side also transitions to turbulence, causing the transverse force to reverse direction, thereby known as reverse swing \cite{Deshpande2018}. A similar reversal in the direction of the lift force is also observed for spinning spheres, commonly referred to as the inverse Magnus effect \cite{Kim2014, Krishnan2025}. In contrast to a smooth sphere, where LSB forms relatively uniformly across the polar region of both hemispheres \cite{Deshpande2017, Achenbach1972}, a cricket ball with asymmetric seam develops an LSB and a secondary vortex (SV) only on the seam side during conventional swing \cite{Parekh2024, Milner2025}. As $Re$ increases toward the reverse-swing regime, the LSB on the perturbed side disappears, while an LSB begins to form on the unperturbed side. These studies collectively demonstrate that asymmetric surface perturbations can fundamentally reorganize the three-dimensional separation topology and wake structure of a sphere, generating substantial transverse forces even in the absence of body rotation.



Motivated by these observations, Sudarsana et al. \cite{sudarsana2024} systematically investigated the effect of prescribed asymmetric surface roughness using localized dimples distributed over one hemisphere of a sphere. Their experiments demonstrated that lift coefficients of up to $C_L\sim0.4$ can be generated on a non-rotating sphere, comparable to the peak Magnus lift produced by body rotation \cite{Beratlis2012, Muto2012, Kim2014, Krishnan2025}. Remarkably, the drag coefficient remained nearly unchanged across all dimple depth ratios investigated. This behavior contrasts sharply with the classical Magnus effect, where both lift and drag increase with increasing rotation ratio \cite{Beratlis2012, Muto2012, Kim2014, Krishnan2025}. The persistence of nearly constant drag despite substantial lift generation suggests that asymmetric roughness induces a fundamentally different flow mechanism than rotation-induced Magnus lift. Rather than simply delaying separation globally, the asymmetric roughness appears to reorganize the wake asymmetrically while preserving the overall pressure drag.
Despite these findings, the underlying three-dimensional flow physics responsible for lift generation on an asymmetrically roughened sphere remain poorly understood. In addition, the prior experiments~\cite{sudarsana2024} employed two-dimensional particle image velocimetry (PIV) measurements performed in the equatorial plane captured only a single separation location along the measurement plane. However, owing to the asymmetric roughness distribution, the separation topology is expected to vary substantially in the azimuthal direction, resulting in inherently three-dimensional and spatially non-uniform separation dynamics. Consequently, the three-dimensional separation behavior and the near-wake processes responsible for sustaining the asymmetric wake remain unresolved. In particular, it is unclear how the rough--smooth interface modifies the local separated shear layers, how the separated flow from the two hemispheres interacts after detachment, and how these interactions generate substantial lift without a comparable change in mean drag. Furthermore, although the experiments revealed the global force response and mean wake deflection, they could not resolve the near-wall and wake flow structures necessary to establish the detailed mechanisms linking localized asymmetric roughness to transverse force generation.

Addressing these questions requires a fully three-dimensional and time-resolved characterization of the near-wall and wake flow field. The present study therefore employs wall-resolved large eddy simulation (LES) to investigate the flow over a sphere with asymmetric surface roughness. Particular attention is given to the spatial variation of the separation topology, development of turbulent separated shear layers, formation of coherent vortical structures, and their role in coupling the asymmetric surface perturbation to wake deflection and force generation. The forthcoming results reveal a highly complex and intrinsically three-dimensional sequence of coupled flow phenomena linking localized asymmetric roughness to global wake reorganization and lift production. The remainder of the paper is organized as follows. Section~\ref{sec2} describes the numerical methodology and simulation setup. Section~\ref{sec3} presents the force characteristics, surface-flow topology, and wake dynamics associated with the asymmetrically dimpled sphere. Finally, Section~\ref{sec4} summarizes the principal findings and conclusions of the study.
\section{Methodology}\label{sec2}

\subsection{\label{sec2sub1}Governing equations and numerical methods}
The present study employs Large Eddy Simulation (LES), in which the spatially filtered incompressible Navier--Stokes equations for a Newtonian fluid are written as:
\begin{align}
\label{ns1}
\frac{\partial \overline{u}_i}{\partial x_i} & = 0 \\ \nonumber
\frac{\partial \overline{u}_i}{\partial t}+\frac{\partial (\overline{u}_i \overline{u}_j)}{\partial x_j} & = -\frac{1}{\rho}\frac{\partial \overline{p}}{\partial x_i} + \nu\frac{\partial}{\partial x_j}\left(\frac{\partial \overline{u}_i}{\partial x_j} \right) - \frac{\partial \tau_{ij}^r}{\partial x_j},
\end{align}
where $\overline{u}_i$ is the filtered velocity, $\overline{p}$ is the filtered pressure, and $\nu$ is the kinematic viscosity. Here, $\overline{\,\cdot\,}$ denotes spatial filtering and $\langle\,\cdot\,\rangle$ represents time-averaging. A filtered quantity $\overline{f}_i$ is defined as $\overline{f}_i(x)=\int G(x,x')f_i(x')dx'$ where $G(x,x')$ is the filter kernel. The residual (subgrid scale) stress tensor, $\tau_{ij}^r =\overline{u_i u_j} - \overline{u}_i \overline{u}_j$, requires modeling for closure. The current study employs the Wall-Adapting Local Eddy-viscosity model (WALE) \cite{Nicoud1999}, in which the subgrid scale stress is modeled following the eddy viscosity $\nu_t$ hypothesis, which assumes linear proportionality between the subgrid scale stress and the resolved strain rate $\overline{S}_{ij}$ as:
\begin{align}
    \tau_{ij}^r &= -2\nu_t \overline{S}_{ij} \\
    \overline{S}_{ij} &= \frac{1}{2}\left(\frac{\partial \overline{u}_i}{\partial x_j} + \frac{\partial \overline{u}_j}{\partial x_i} \right).
\end{align}
The WALE model expresses the eddy viscosity as \cite{Nicoud1999, openfoam2406}:
\begin{align}
    \nu_t=(C_w\Delta)^2\frac{(S_{ij}^dS_{ij}^d)^{3/2}}{(\overline{S}_{ij}\overline{S}_{ij})^{5/2}+(S_{ij}^dS_{ij}^d)^{5/4}},
\end{align}
where $S_{ij}^d$ is the traceless, symmetric part of the square velocity gradient tensor, $\Delta=(\Delta_x\Delta_y\Delta_z)^{1/3}$ is the sub-grid characteristic length scale computed from the local mesh spacings, and $C_w$ is a model constant. The value $C_w=0.325$ is used in the current study, following previous studies \cite{qin2018, kim2019, Kim2020}. With the velocity gradient defined as $\overline{g}_{ij}=\partial\overline{u}_i/\partial x_j$ and its square as $\overline{g}_{ij}^2=\overline{g}_{ik}\overline{g}_{kj}$, the traceless symmetric tensor can be expressed as:
\begin{align}
    S_{ij}^d=\frac{1}{2}\left(\overline{g}_{ij}^2 + \overline{g}_{ji}^2 \right) - \frac{1}{3}\delta_{ij}\overline{g}_{kk}^2,
\end{align}
where $\delta_{ij}$ is the Kronecker delta.

The WALE model is selected for the present configuration because the invariant $(S^d_{ij}S^d_{ij})$ depends on both the strain rate and the rotation rate of the resolved field, so the eddy viscosity responds to the rotation and strain interaction produced by the dimples and the separating shear layer. Two further properties matter here. First, the eddy viscosity recovers the correct near-wall asymptotic behavior $\nu_t\sim y^3$ \cite{Nicoud1999, qin2018} without \emph{ad-hoc} damping \cite{Nicoud1999, Arya2019, Kim2020, qin2018}, so the resolved dimple-induced near-wall structures that drive boundary-layer transition and separation are preserved. Second, $S^d_{ij}$ vanishes in regions of pure shear such as a laminar boundary layer, so the model adds no spurious dissipation in laminar zones and does not over-damp the instability modes that initiate transition. Secondary instabilities and subharmonic resonance can therefore develop on the resolved grid through natural amplification rather than being suppressed by the subgrid scale model, as demonstrated in previous LES of transitional flows \cite{Kim2020, qin2018}.

The governing equations are solved using an open-source solver, \texttt{OpenFOAM v2406}. Pressure-velocity coupling is handled with the PIMPLE algorithm, which combines the Pressure Implicit with Splitting of Operators (PISO) algorithm~\cite{Issa1986} with the Semi-Implicit Method for Pressure Linked Equations (SIMPLE)~\cite{Patankar1972}. A second-order accurate implicit Backward method is employed for time integration. The spatial derivatives are discretized using the finite-volume method. The convective term, $\frac{\partial (\overline{u}_i \overline{u}_j)}{\partial x_j}$, is evaluated using a bounded Gauss linear upwind scheme, which provides second-order accuracy. Diffusion terms are discretized using Gauss linear discretization, and Laplacian operators use Gauss linear corrected, which includes non-orthogonality corrections for unstructured meshes. Face interpolation of cell-centered variables uses linear interpolation, and face-normal gradients use a limited formulation that blends corrected and uncorrected contributions with a weighting of 0.5. The momentum predictor is solved with a Preconditioned
Bi-Conjugate Gradient (PBiCG) iterative method preconditioned by Diagonal Incomplete-Lower-Upper (DILU), while the pressure correction uses Preconditioned Conjugate Gradient (PCG) with Diagonal Incomplete Cholesky (DIC) preconditioning. For both, a tolerance of $10^{-6}$ is used. The CFL (Courant–Friedrichs–Lewy) number is maintained below 0.4. More details of the numerical method are available in ~\cite{openfoam2406}.

\subsection{\label{sec2sub2}Problem set-up}
\begin{figure}[htbp]
    \centering
    \includegraphics[width=0.85\linewidth]{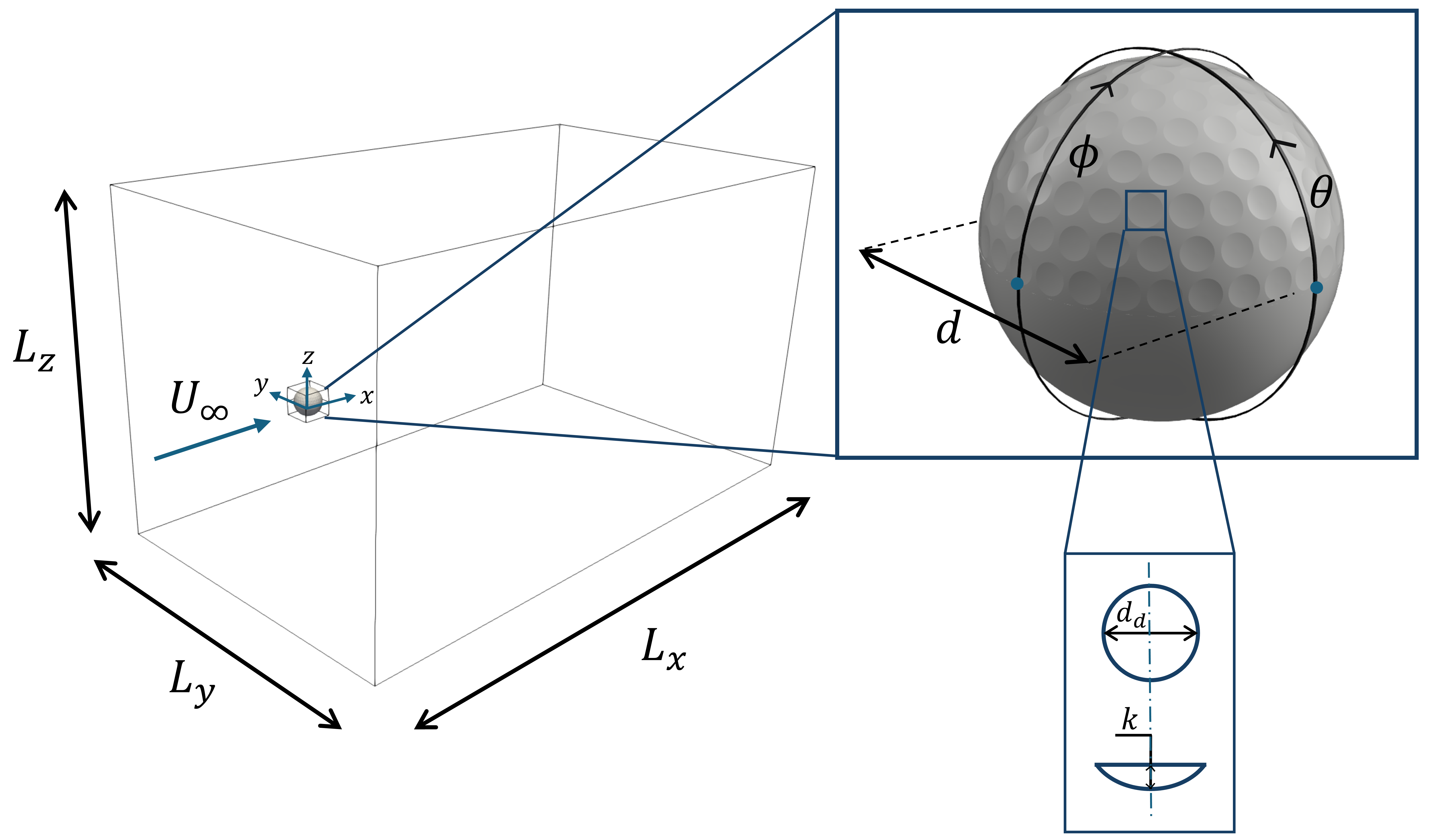}
    \caption{Computational domain and sphere with an asymmetric dimple configuration. The geometric parameters are shown on the right. The domain dimensions are denoted by $L_x, L_y, L_z$, while the sphere diameter, dimple depth, and dimple diameter are denoted by $d$, $k$ and $d_d$, respectively. The freestream velocity, $U_{\infty}$, is directed from left to right. The spherical-coordinate system used to define the polar angle, $\theta$, and azimuthal angle, $\phi$, is also shown in the model schematic on the right, where the blue dots indicate the reference locations corresponding to $\theta=0$ and $\phi=0$, along the $-y$ and $-x$ directions, respectively.}
    \label{fig:dom-geom}
\end{figure}

Four configurations are considered: a smooth-surface sphere and three asymmetrically roughened sphere with varying dimple depths in which one hemisphere is dimpled and the other hemisphere remains smooth. These configurations are simulated at $Re=100{,}000$, which lies within the sub-critical regime for a smooth sphere \cite{sudarsana2024,Achenbach1972}.  Three dimple depth-to-sphere-diameter ratios of $k/d=0.004$ (Fig.~\ref{fig:dom-geom}), $k/d=0.006$, and $k/d=0.008$ are considered. These dimple configurations follow previous experimental studies \cite{sudarsana2024, vilumbrales2025}, with a dimple area coverage ratio $AR= 55.6\%$, computed from $AR = N_d d_d^2/(2d^2)$, where $N_d = 148$ is the total number of dimples on the roughened hemisphere and $d_d/d = 0.087$ is the dimple-to-sphere diameter ratio, comparable to that used in \cite{Choi2006}. The configuration and parameter space are chosen to match the experimental conditions under which transverse (lift) force generation was observed in \cite{sudarsana2024}. While the optimal case giving the largest lift was $k/d=0.003$ in the experiments, the current study chooses $k/d>0.003$ to compensate for the effect of freestream turbulence ($TI\approx1.8\%$) in the experimental setup \cite{sudarsana2024}. No freestream turbulence is imposed in the current simulations.

The computational domain is a rectangular box of dimensions ($L_x, L_y, L_z$) = ($24.3d$, $13.5d$, $13.5d$), where $x$ represents the streamwise direction, $y$ and $z$ are the stream-normal directions, as shown in Fig.~\ref{fig:dom-geom}. This domain size is comparable to that used in previous LES studies with box domains~\cite{Muto2012, Krishnan2025} and DNS studies with cylindrical domains~\cite{Beratlis2012, Beratlis2019} at similar Reynolds numbers. The sphere center is located at $3d$ from the inlet or $21.3d$ from the outlet (along the $x$-coordinate) and $6.75d$ from each side wall (along $y$ and $z$ coordinates). A uniform freestream velocity is imposed at the inlet as a Dirichlet boundary condition with a Neumann condition for the pressure. At the outlet, the pressure is fixed via a Dirichlet condition, and the velocity satisfies a zero-gradient (Neumann) condition. Slip conditions are applied on the side walls together with a Neumann condition on the pressure, consistent with previous numerical studies of flow over spheres \cite{Beratlis2019, Parekh2024, Beratlis2012, Krishnan2025}. A no-slip velocity condition is imposed on the sphere surface, and the turbulent viscosity is set to zero ($\nu_t=0$), consistent with wall-resolved LES. 


The computational domain is discretized into finite volume hexahedral cells using \texttt{blockMesh} and \texttt{snappyHexMesh}. Two mesh configurations are employed: approximately 44 million cells for the smooth sphere and $53-60$ million cells for the asymmetric dimpled spheres for different $k/d$. Both mesh configurations use five levels of successive refinement, each level subdividing cells by a factor of $2^n$ from the base mesh, with $n$ is the level number. The near-wake box refinement extends to level 3 (a cell size of $0.01d$), while a spherical refinement around the sphere ($0.1d$ from the sphere's surface) reaches level 5 or a cell size of $0.0026d$. A surface layer is added to the near-wall region with a first-layer height of approximately $0.0001d$ for $k/d=0.004$ and $0.00005d$ for $k/d=0.006$ and $k/d=0.008$. This yields a near-wall resolution of $y^+<1$, consistent with an earlier study of the wall-resolved LES of flow over sphere \cite{Parekh2024} and the well-accepted guidelines of wall-resolved LES. Here $y^+$ is defined in viscous units as $ y^+ = r u_\tau/\nu$ with $u_\tau =\sqrt{\tau_w/\rho}$, $\tau_w=\mu \frac{\partial u_t}{\partial n}\Bigr|_{\substack{n=0}}$, $n$ the wall-normal direction, $u_t$ the tangential velocity, and $r$ the distance from the sphere surface. The streamwise and spanwise grid spacings in wall units are approximately $\Delta x^+=\Delta z^+= 0.12-8.1$. The smooth and asymmetric-dimpled sphere meshes differ primarily in the wake-refinement region, where the asymmetric-dimpled configuration employs a larger refined volume downstream of the sphere than the smooth case to capture the wake behavior. This wake refinement strategy for the asymmetric-dimpled sphere is selected \emph{a posteriori} based on the observed wake deflection. The simulations were run using 1440 CPU cores for smooth and $k/d=0.004$ cases and 1512 CPU cores for $k/d=0.006$ and $k/d=0.008$. Simulations were first run for $\approx40$ hours of wall time on a coarser mesh, and then for $\approx50-60$ hours on a finer mesh to reach a dynamical steady state. After that, the statistics were collected over $\approx90-120$ hours. In total, the smooth sphere took $\approx250$ thousand CPU core hours while the dimpled case took $\approx300-340$ thousand CPU core hours.

\begin{figure}[htbp]
    \centering
    \includegraphics[width=0.7\linewidth]{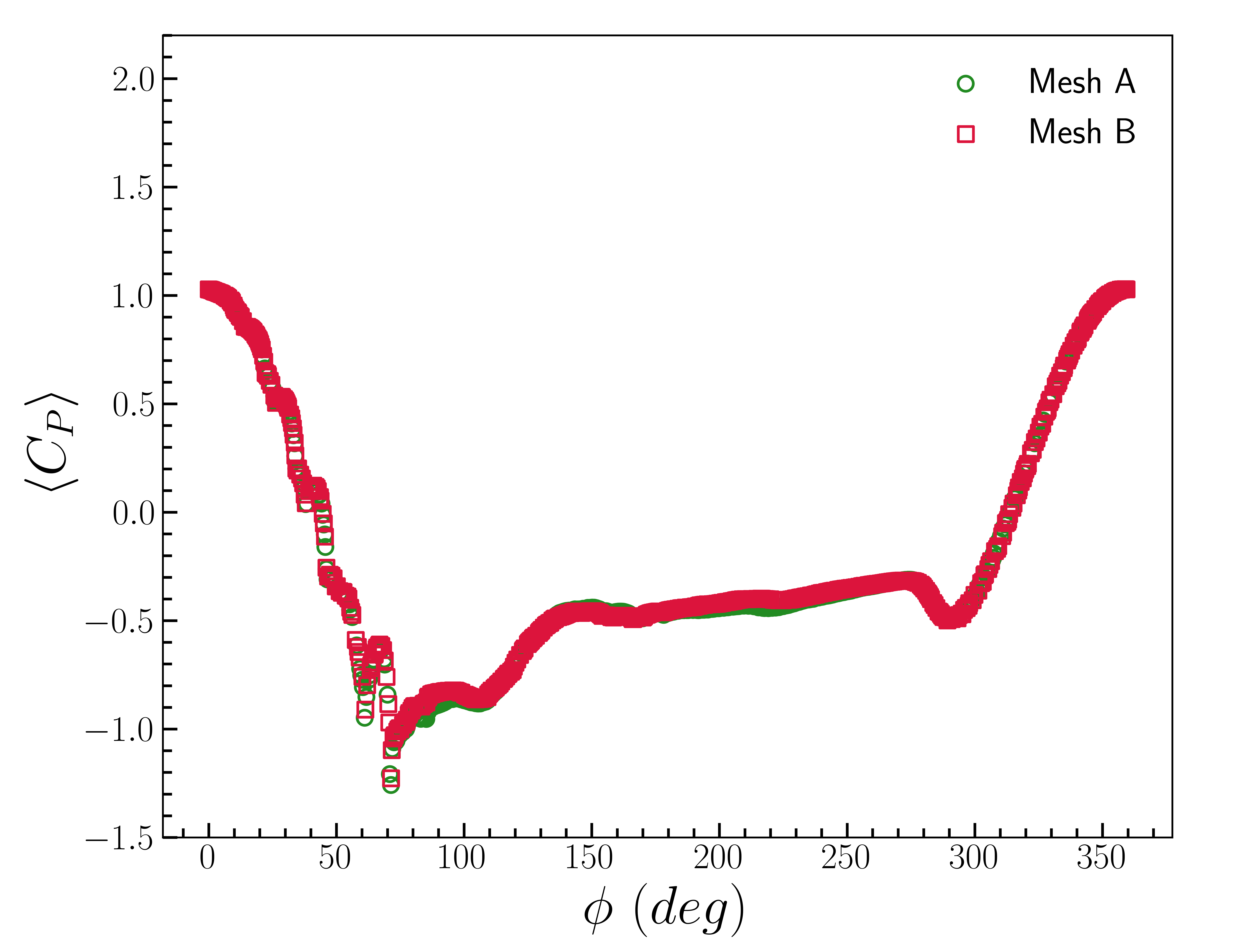}
    \caption{Comparison of the time-averaged surface pressure coefficient, $\langle C_P \rangle$ as a function of azimuthal angle $\phi$ (in degrees) between mesh-A (31.5 million cells) and mesh-B (53 million cells) for the asymmetrically dimpled sphere with $k/d=0.004$.}
    \label{fig:cp_mesh}
\end{figure}

The mesh convergence analysis is conducted on the sphere with asymmetric dimples, by evaluating the time-averaged force coefficients and the time-averaged surface pressure $\langle C_P \rangle$ in Fig. \ref{fig:cp_mesh}. The asymmetric dimpled case with $k/d=0.004$ was run with two mesh configurations: Mesh-A and Mesh-B, which have a total mesh count of 31.5 million and 53 million cells, respectively. The main difference between the meshes is in the refinement of the wake regime, where Mesh-B has finer mesh resolution (two times smaller mesh size) compared to Mesh-A. Mesh-A and Mesh-B shows a small difference (less than 3\%) in the time-averaged drag coefficient $\langle C_D \rangle= 2\langle F_D\rangle/(\rho U_{\infty}^2 A)$ and lift coefficient $\langle C_L \rangle= 2\langle F_L\rangle/(\rho U_{\infty}^2 A)$, where $A=\pi d^2/4$. Mesh-A exhibits slightly larger values of both $\langle C_D\rangle$ and $\langle C_L\rangle$, with values of 0.542 and 0.331, respectively, compared with 0.521 and 0.325 for Mesh-B. Further check of the time-averaged pressure coefficient, $\langle C_P \rangle=2\langle p\rangle/(\rho U_{\infty}^2)$ at the $x$-$z$ mid-plane also shows no significant variations between the predictions using Mesh-A and Mesh-B (Fig.~\ref{fig:cp_mesh}). This confirms the adequacy of the mesh-B for the current analysis. All of the analysis in this study is conducted using Mesh-B, to resolve the vortical structures in the wake region.

\subsection{\label{sec2sub3}Verification and validation}
The simulation framework is verified by comparing the current results of the smooth sphere with those available in the literature at similar Reynolds numbers. Table~\ref{tab:force_comp} summarizes the time-averaged drag coefficients $\langle C_D \rangle$ and lift coefficients $\langle C_L \rangle$ of the smooth sphere at $Re=100{,}000$. The present LES results are in good agreement with several experimental studies \cite{Achenbach1972, Kim2014, sudarsana2024, vilumbrales2025}, with 2.1\% lower compared to Achenbach (1972) \cite{Achenbach1972} and Kim et al. (2014) \cite{Kim2014} and 4.5\% lower compared to Sudarsana et al. (2024) \cite{sudarsana2024} in $\langle C_D \rangle$. A large difference in the $\langle C_D \rangle$ with Vilumbrales-Garcia et al. \cite{vilumbrales2025} ($\sim10\%$) can be attributed to the higher freestream turbulent intensity, $TI$, reported in the study while the current simulations did not introduce any freestream perturbations. The current prediction of $\langle C_D \rangle$ is also within 3.2\% of the value reported in the LES study by Krishnan et al. \cite{Krishnan2025}. As expected, the time-averaged lift coefficient is $\langle C_L \rangle \approx 0$ in all studies.


\begin{table*}[htbp]
\caption{\label{tab:force_comp}Validation of the smooth sphere result with previous studies at $Re=100{,}000$}
\begin{tabular}{lcc}
\hline
      ~~~~~~~~ &  $\langle C_D \rangle$ &  $\langle C_L \rangle$  \\
      \hline
       Present study (LES)  & 0.505 & -0.0086 \\
       Krishnan et al. \cite{Krishnan2025} (LES)  & 0.489$\pm$0.0071 &  0.0038  \\
       Sudarsana et al. \cite{sudarsana2024} (Exp.)  & 0.529$\pm$0.015 &  0.021$\pm$0.012  \\
       Vilumbrales-Garcia et al. \cite{vilumbrales2025} (Exp.)  & 0.56$\pm$0.015 &  -  \\
       Achenbach et al. \cite{Achenbach1972} (Exp.)  & 0.516 & -  \\
       Kim et al. \cite{Kim2014} (Exp.) & 0.516 & -0.0052 \\
       \hline
\end{tabular}
\end{table*}

\begin{figure}[htbp]
    \centering
    \includegraphics[width=0.7\linewidth]{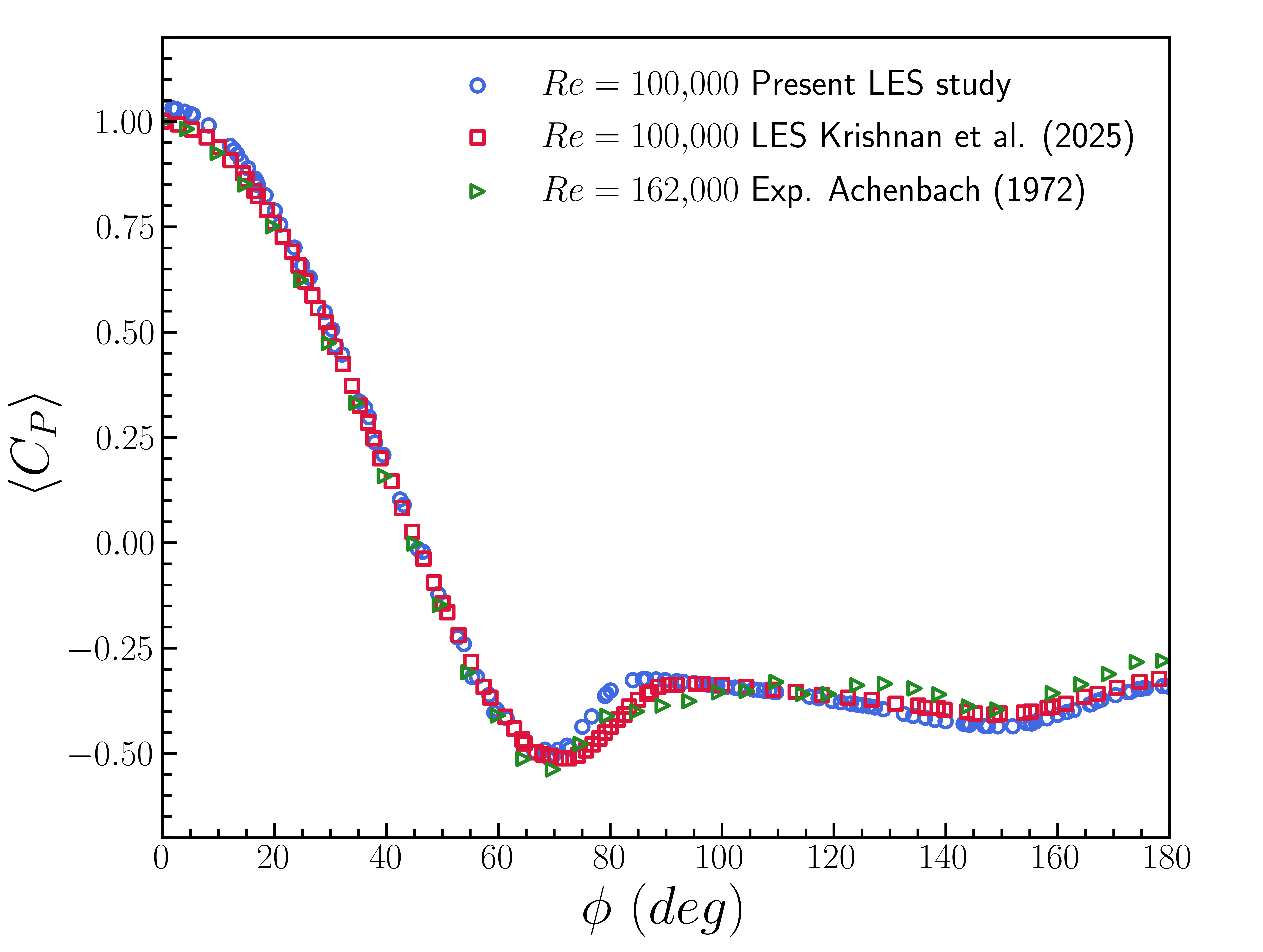}
    \caption{Comparison of the time-averaged surface pressure coefficient, $\langle C_P \rangle$ with previous experimental \cite{Achenbach1972} and simulation study \cite{Krishnan2025} for the smooth sphere.}
    \label{fig:cp_compare}
\end{figure}

Further validation beyond the integral parameters is also considered by comparing the time-averaged surface pressure coefficient $\langle C_P\rangle$ and the streamwise velocity profile $\langle u_x\rangle/U_{\infty}$ with previous studies. As seen in Fig~\ref{fig:cp_compare}, the current LES predictions of  $\langle C_P\rangle$ at $Re=100{,}000$ are in good agreement with previous experimental and simulation studies at $Re=162{,}000$ and $Re=100{,}000$, respectively. The overall $\langle C_P\rangle$ shows a good agreement with previous simulation results, with small differences at $\phi\sim80^\circ$ and $\phi\sim150^\circ$~\cite{Krishnan2025}. This is also consistent with the small difference observed in the prediction of $\langle C_D\rangle$. This difference can be attributed to the different LES models used and to a slight difference in the mesh. The current LES at $Re=100{,}000$ also shows a surface pressure $\langle C_P\rangle$ profile consistent with the experimental measurements at $Re=162{,}000$ \cite{Achenbach1972}, with minor discrepancies at $\phi \geq 80^\circ$, which can be attributed to the different values of $Re$. This comparison is valid because both $Re$ lie within the subcritical regime, where $C_D$ remains largely independent of the $Re$ and constant at $\approx0.5$ \cite{Achenbach1972, Deshpande2017, vilumbrales2025}.

\begin{figure}[htbp]
    \centering
    \includegraphics[width=1\linewidth]{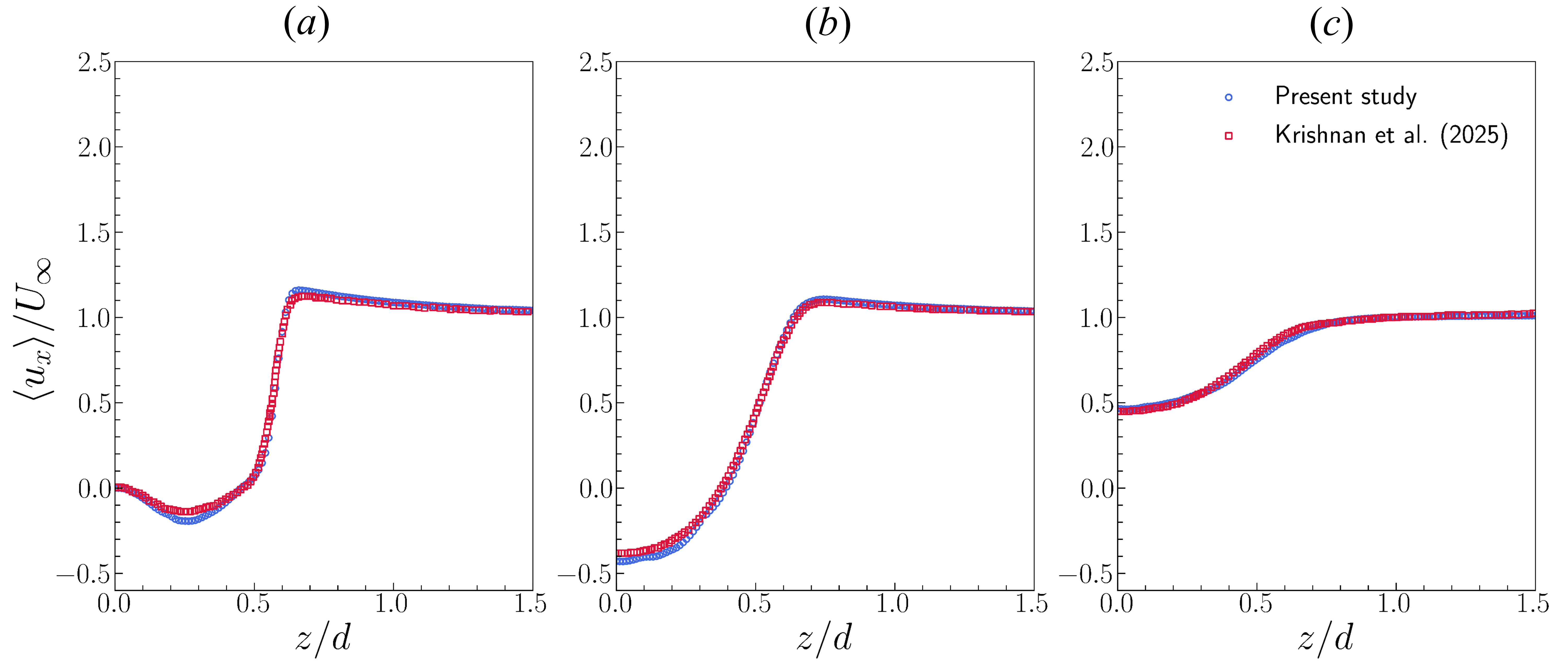}
    \caption{Comparison of the time-averaged normalized streamwise velocity, $\langle u_x \rangle /U_{\infty}$, as a function of $z/d$ at (a) $x/d=0.5$, (b) $x/d=1$, and (c) $x/d=1.5$, with the previous simulation results of Krishnan et al. \cite{Krishnan2025} for the smooth sphere.}
    \label{fig:linewake_compare}
\end{figure}

Figure~\ref{fig:linewake_compare} shows the comparison of the time-averaged normalized streamwise velocity $\langle u_x \rangle /U_\infty$ in the near-wake region at three different $x/d$ locations. Again, an excellent agreement is shown between the current LES results with the WALE model and the LES results of Krishnan et al. \cite{Krishnan2025} with $k$-equation model, where the subgrid scale kinetic energy is solved via transport equation to compute the eddy viscosity $\nu_t$. 



\section{Results and Discussion}\label{sec3}
The main aim of this study is to investigate the effect of asymmetric dimples on wake characteristics in case of the flow past a sphere to explain the mechanism of transverse force generation observed in our previous experimental study \cite{sudarsana2024}. The analysis begins in a broad point-of-view by analyzing the forces and surface flow behavior, which will be followed by near-wall and near-wake flow characteristics, focusing on the onset of asymmetric wake.

\subsection{\label{sec3sub1} Integral force-coefficients, surface pressure distribution and flow separation}

 Table \ref{tab:force_comp} compares the time-averaged drag coefficient $\langle C_D \rangle$ and lift coefficient $\langle C_L \rangle$ for the smooth and asymmetric dimpled spheres from the present LES with the experimental measurements~\cite{sudarsana2024}. Statistics were collected for $\Delta t^* \approx 120$ for all cases after reaching a dynamical steady state. Here, a time unit $t^*=tU/d=1$ corresponds to one flow through time past the sphere. This averaging window is approximately four times larger than the previous LES study \cite{Parekh2024} and is comparable to the averaging window used in the DNS study of dimpled spheres \cite{Beratlis2019}. For the smooth sphere, mean drag, $\langle C_D \rangle$, predictions from the LES agree well with the experiments and, as expected, $\langle C_L \rangle \approx 0$ in both experiments~\cite{sudarsana2024} and current simulations. In contrast to the smooth sphere, the mean lift force is non-zero for the asymmetrically dimpled sphere, with the LES under-predicting $\langle C_L \rangle$ by 10\% compared to the experiments. A small difference in the predictions of $\langle C_D \rangle$ is also evident for the asymmetrically dimpled spheres between LES and experiments. This difference can be attributed to the lack of freestream turbulence in the LES. The freestream turbulence levels in the experiments were of the order of 1.8\%, which can further delay the flow separation on the dimpled side, leading to slightly higher values of $\langle C_L \rangle$ compared to the LES with no freestream turbulence. Table \ref{tab:force_comp1} shows that the pressure drag contributes predominantly to the total drag, about 98\%. For the three dimple depths investigated, the mean lift coefficient $\langle C_L \rangle$ varies by $\approx 11\%$ while the mean drag coefficient $\langle C_D \rangle$ varies only by $\approx2\%$. The mean lift coefficient $\langle C_L \rangle$ first increases with increasing $k/d$ before it starts decreasing, which is consistent with the experimental observations~\cite{sudarsana2024}. The mechanism for the lift generation is discussed in Sec.~\ref{sec3sub4}. Overall, the current LES results are consistent with the past experimental study under similar conditions, and show that even in the absence of freestream turbulence intensity, dimples placed asymmetrically on the sphere surface can lead to lift (transverse force) generation that is comparable or higher than the maximum achievable lift during the classical Magnus effect \cite{Kim2014}. 

\begin{table*}[htbp]
\caption{\label{tab:force_comp}Comparison between LES (present study) and experimental result \cite{sudarsana2024} for the smooth and asymmetrically dimpled sphere at $Re=100{,}000$.}
\begin{tabular}{lcccc}
\hline
      ~~~~~~~~ &  \multicolumn{2}{c}{LES} & \multicolumn{2}{c}{Experiment \cite{sudarsana2024}}  \\
      ~~~~~~~~ &  $\langle C_D \rangle$ &  $\langle C_L \rangle$ &  $\langle C_D \rangle$ &  $\langle C_L \rangle$  \\
      \hline
       Smooth & 0.505 & -0.0086 & 0.529$\pm$0.015 &  0.021$\pm$0.012  \\
       Asymmetric dimpled $k/d=0.004$ & 0.521 & 0.322 & 0.515$\pm$0.015 &  0.38$\pm$0.02 \\
       Asymmetric dimpled $k/d=0.006$ & 0.505 & 0.359 & 0.512$\pm$0.012 & 0.366 $\pm$0.011 \\
       Asymmetric dimpled $k/d=0.008$ & 0.509 & 0.320 & 0.512$\pm$0.012 & 0.34 $\pm$0.014 \\
       \hline
\end{tabular}
\end{table*}

\begin{table*}[htbp]
\caption{\label{tab:force_comp1}Comparison between the smooth and asymmetrically dimpled sphere for different $k/d$ at $Re=100{,}000$.}
\begin{tabular}{lcccc}
\hline
      ~~~~~~~~ &  $\langle C_D \rangle$ &  $\langle C_L \rangle$ & $\langle C_{D_p} \rangle/\langle C_D \rangle $ & $\langle C_{L} \rangle/\langle C_D \rangle $ \\
      \hline
       Smooth, LES & 0.505 & -0.0086 & 98\% & $\approx0$  \\
       Asymmetric dimpled $k/d=0.004$, LES & 0.521 & 0.322 & 97.79\% & 0.618 \\
       Asymmetric dimpled $k/d=0.006$, LES & 0.505 & 0.359 & 97.66\% & 0.711 \\
       Asymmetric dimpled $k/d=0.008$, LES & 0.509 & 0.320 & 97.42\% & 0.628 \\
       \hline
\end{tabular}
\end{table*}

\begin{figure}[htbp]
    \centering
    \includegraphics[width=0.9\linewidth]{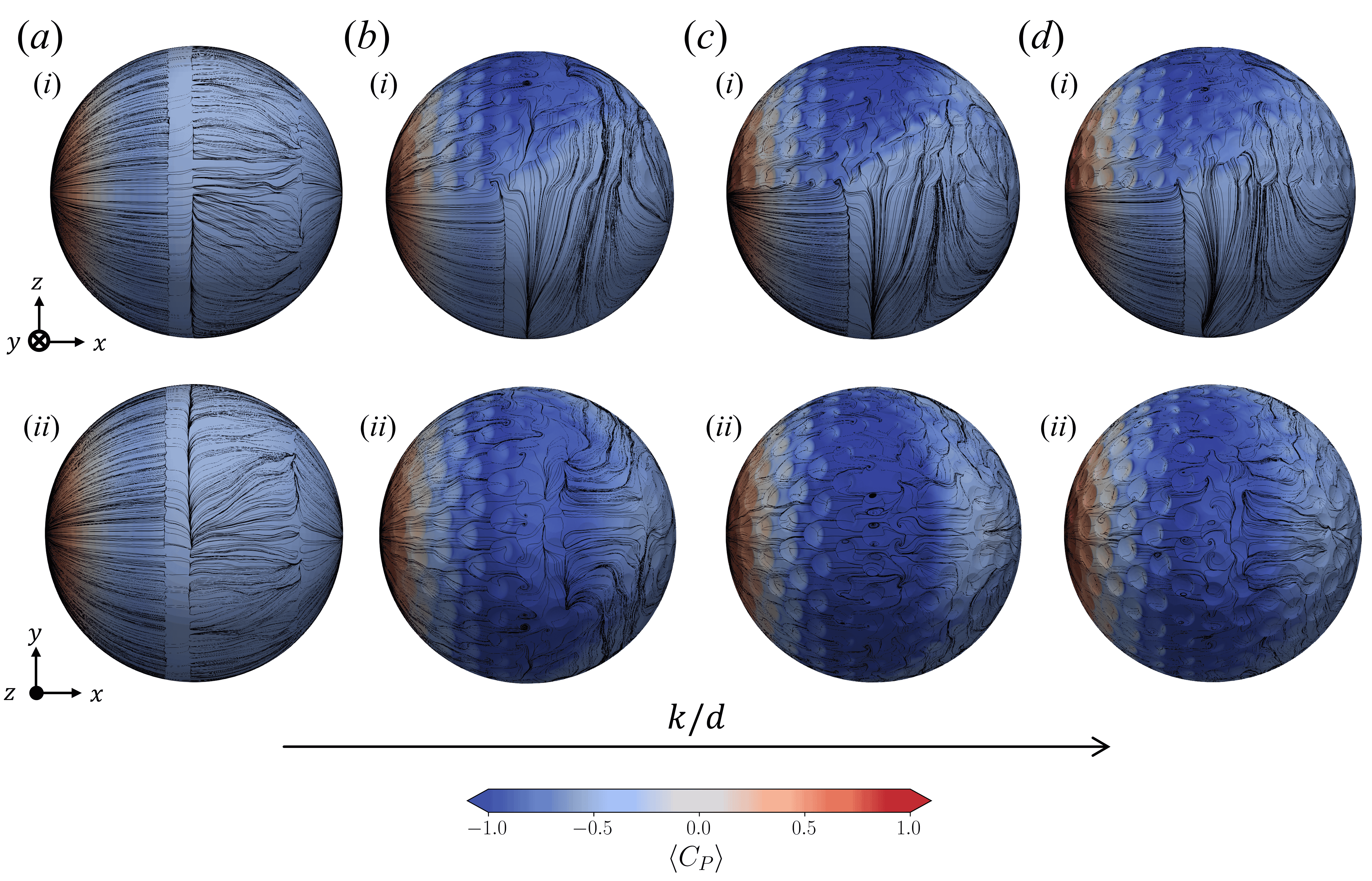}
    \caption{Time-averaged surface pressure coefficient $\langle C_P\rangle$ for the (a) Smooth sphere and  asymmetrically dimpled spheres with (b) $k/d=0.004$, (c) $k/d=0.006$, and (d) $k/d=0.008$, overlaid with the time-averaged surface streamlines. Panels (i) and (ii) correspond to the side and the top view, respectively.}
    \label{fig:surfaceplot}
\end{figure}

\begin{figure}[htbp]
    \centering
    \includegraphics[width=0.6\linewidth]{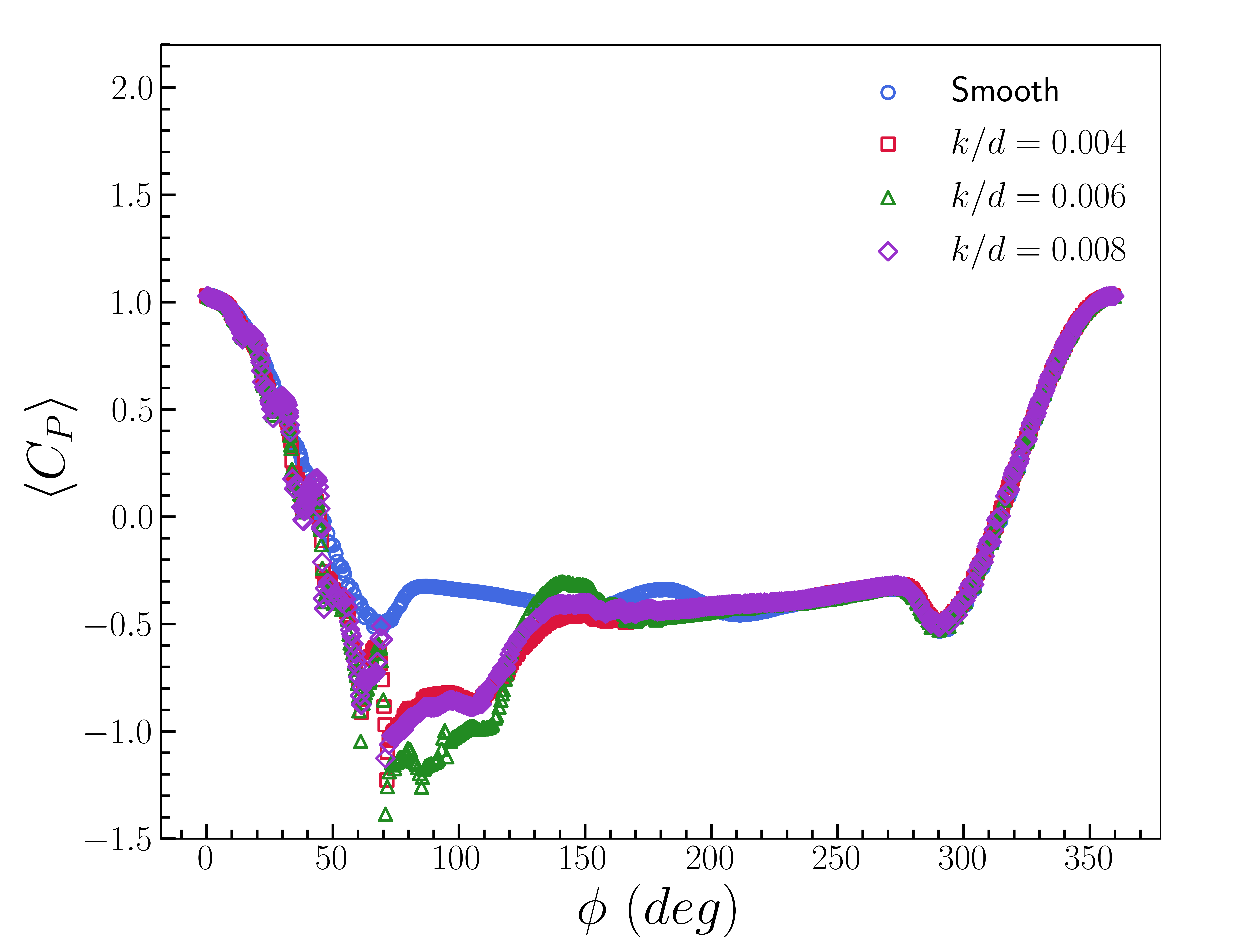}
    \caption{Time-averaged surface pressure coefficient $\langle C_P\rangle$ along the midplane ($\theta=90^\circ$) for the smooth sphere and asymmetrically dimpled spheres at various dimple depth ratios.}
    \label{fig:cpsmodim}
\end{figure}

The effect of asymmetry of the surface roughness due to dimples on the pressure distribution is investigated in Fig.~\ref{fig:surfaceplot} which shows time-averaged surface streamlines (skin friction lines) overlaid on the time-average pressure coefficient, $\langle C_P \rangle$, contours. The time-averaged surface streamlines are plotted by generating vector lines that follow the direction of the surface wall shear stress $\tau_w$ vector fields, where $\tau_w=\tau \cdot \mathbf{n}$ with $\mathbf{n}$ denoting the surface normal pointing outward. The surface $\langle C_P \rangle$ behaves as expected on the smooth sphere (Fig.~\ref{fig:surfaceplot}a), where the $\langle C_P \rangle$ distribution is nearly axisymmetric for a given $x$ and the asymmetry in $\langle C_P \rangle$ between the front and the rear sides contributes to the total drag. This is also reflected in the surface $\langle C_P \rangle$ profile at $\theta=90^\circ$ (midplane passing through the sphere center) over different azimuthal $\phi$ angles (Fig.~\ref{fig:cpsmodim}). The symmetry of $\langle C_P \rangle$ between the top ($\phi=0^\circ-180^\circ$) and bottom sides ($\phi=180^\circ-360^\circ$), reflects the absence of lift generation $\langle C_L \rangle\approx0$ in the smooth sphere case. The addition of dimples on the half hemisphere strongly modifies the $\langle C_P \rangle$ distribution, with significantly lower values of $\langle C_P \rangle$ on the dimpled side indicating higher suction on the dimpled side compared to the smooth side (Fig.~\ref{fig:surfaceplot}b). This is also reflected in the asymmetry of $\langle C_P \rangle$-profile between the top and bottom surfaces over the midplane $\theta=90^\circ$ (Fig.~\ref{fig:cpsmodim}). This asymmetry in the surface $\langle C_P \rangle$ manifests as the generation of lift force. The maximum suction on the front side is observed for $k/d=0.006$ for which the maximum lift force coefficient and the minimum drag force coefficient were observed in Table~\ref{tab:force_comp1}.

\begin{figure}[htbp]
    \centering
    \includegraphics[width=1\linewidth]{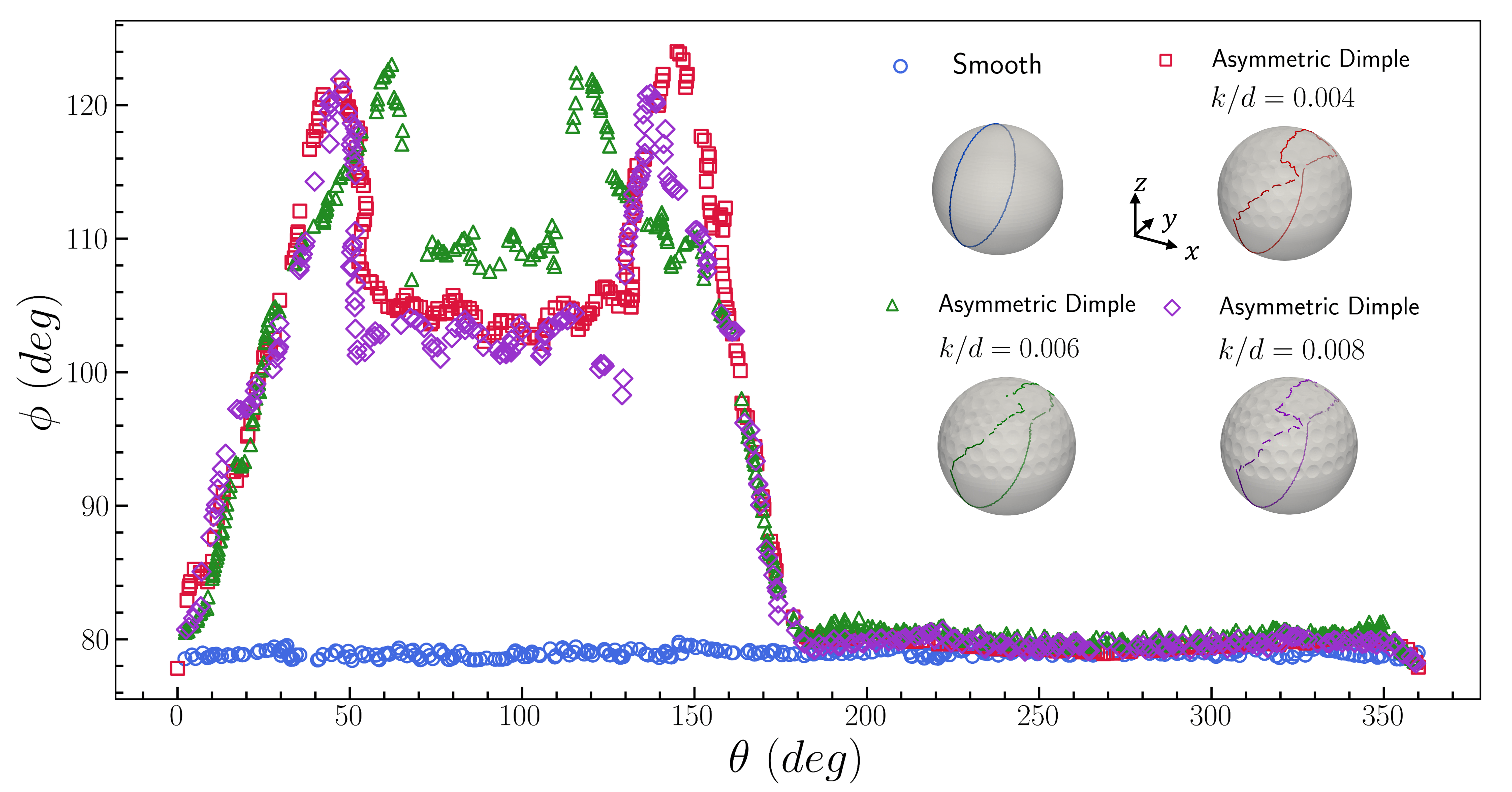}
    \caption{Global separation angle estimated using the condition $\langle {C_f}_x\rangle \lesssim 0$ together with the surface skin-friction topology shown in Figure~\ref{fig:surfaceplot}. Separation locations are identified from the convergence of skin friction lines following the criteria of \cite{lighthill1963attachment} and Surana et al. \cite{Surana2006}. The sphere schematics in the legend are depicted with an opacity of 0.5 for clarity. }
    \label{fig:3dsepsmodim}
\end{figure}

\begin{figure}[htbp]
    \centering
    \includegraphics[width=1\linewidth]{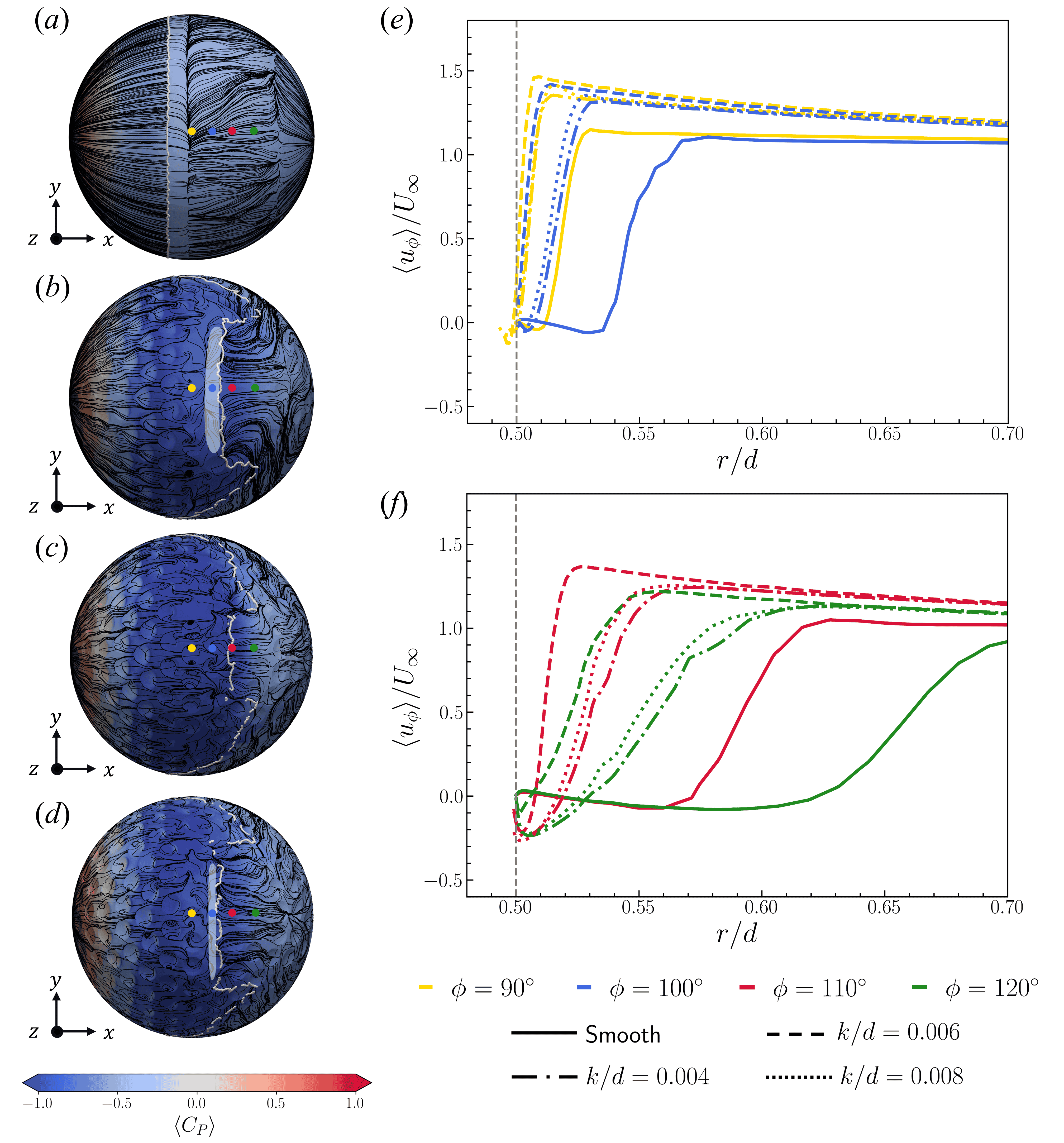}
    \caption{Time-averaged surface pressure coefficient, $\langle C_p \rangle$, overlaid with skin-friction lines on the asymmetric dimpled sphere for (a) $k/d=0.000$ (smooth), (b) $k/d=0.004$, and (c) $k/d=0.006$, and (d) = $k/d=0.008$ and corresponding tangential velocity profiles, $\langle u_\phi \rangle/U_\infty$ (shown on the right column), at (e) $\phi=90^\circ$ and $100^\circ$ and (f) $\phi=110^\circ$ and $120^\circ$ along $\theta=90^\circ$. The gray dashed line  in (a)-(d) denotes the mean separation location, and the white shaded region marks the separated region where the secondary vortex pair forms. Colored dots on the sphere indicate the azimuthal locations at which the tangential velocity profiles are extracted. In panels (e,f), the vertical gray dashed line at $r/d=0.5$ denotes the sphere surface.}
    \label{fig:cpvelprofile}
\end{figure}

The asymmetry in $\langle C_P \rangle$ profile with asymmetric dimples, as seen in Fig.~\ref{fig:cpsmodim}, does not represent the overall pressure distribution on the sphere surface, given its three-dimensionality; in this case, it merely gives a snippet of the asymmetry in $\langle C_P \rangle$ over $\phi$ at $\theta=90^\circ$, therefore, cannot be used reliably to predict the location of flow separation. To identify the flow separation locations over the sphere surface, Fig.~\ref{fig:3dsepsmodim} plots the global separation location ($\phi$) over polar angle $\theta$ based on $\langle {C_f}_x\rangle \lesssim 0$, and convergence of skin friction lines \cite{lighthill1963attachment} along with other separation pattern criteria as outlined by Surana et al.~\cite{Surana2006}. The separation line pattern criteria are characterized by the topological connections of the skin-friction field on the boundary, with four different patterns comprising saddle-spiral connections, saddle-node connections, saddle-limit cycle connections, and purely open limit cycles \cite{Surana2006}. This flow separation estimation approach is also confirmed with the Moore-Rott-Sears (MRS) criterion \cite{Moore1958, Rott1956UnsteadyVF, Sears1956SomeRD} where flow separation should satisfy two conditions: $u_\phi=0$ and $\partial u_\phi/\partial r=0$. For validation purposes, $\phi_{sep}$ estimation with MRS criterion is only implemented on the midplane ($\theta=90^\circ$), showing a similar $\phi_{sep}$ trend for smooth and different $k/d$ with small discrepancies ($\sim 4^\circ$) compared to the estimation approach in Fig. \ref{fig:3dsepsmodim} (see Appendix \ref{appendix1}). The separation location on a smooth sphere is constant at $\phi\sim80^\circ$, consistent with past experimental studies~\cite{Achenbach1972, Kim2014}. In contrast, the flow on the dimpled hemisphere in the case of an asymmetrically dimpled sphere, exhibits a varying separation angle location as seen for $\phi, \theta\in[0^\circ,180^\circ]$ while on the smooth hemisphere ($\phi, \theta\in[180^\circ,360^\circ]$), the separation location is uniform at $\phi\sim80^\circ$. On the dimpled side  for $k/d=0.004$, the separation location reaches its local peak ($\phi_{sep}\sim120^\circ$) at $\theta\sim50^\circ$ and $\theta\sim150^\circ$ while at the top within $\theta\sim55^\circ-125^\circ$, it remains nearly constant at $\phi\sim105^\circ$. This behavior can be attributed to the regions of high suction that drive the flow from the higher pressure region on the smooth hemisphere side towards the low pressure region on the dimple side, thereby delaying flow separation close to the interface between smooth and dimpled hemispheres. This leads to a smaller region of flow separation on the dimpled hemisphere around the polar region (near $\theta\sim90^\circ$) as shown by the friction lines in Fig.~\ref{fig:cpvelprofile}. This separation region shrinks and the separation location shifts downstream due to stronger sidewash from the smooth side of the sphere to the dimpled side as $k/d$ approaches its optimum value that gives the maximum lift coefficient.


\begin{figure}[htbp]
    \centering
    \includegraphics[width=1\linewidth]{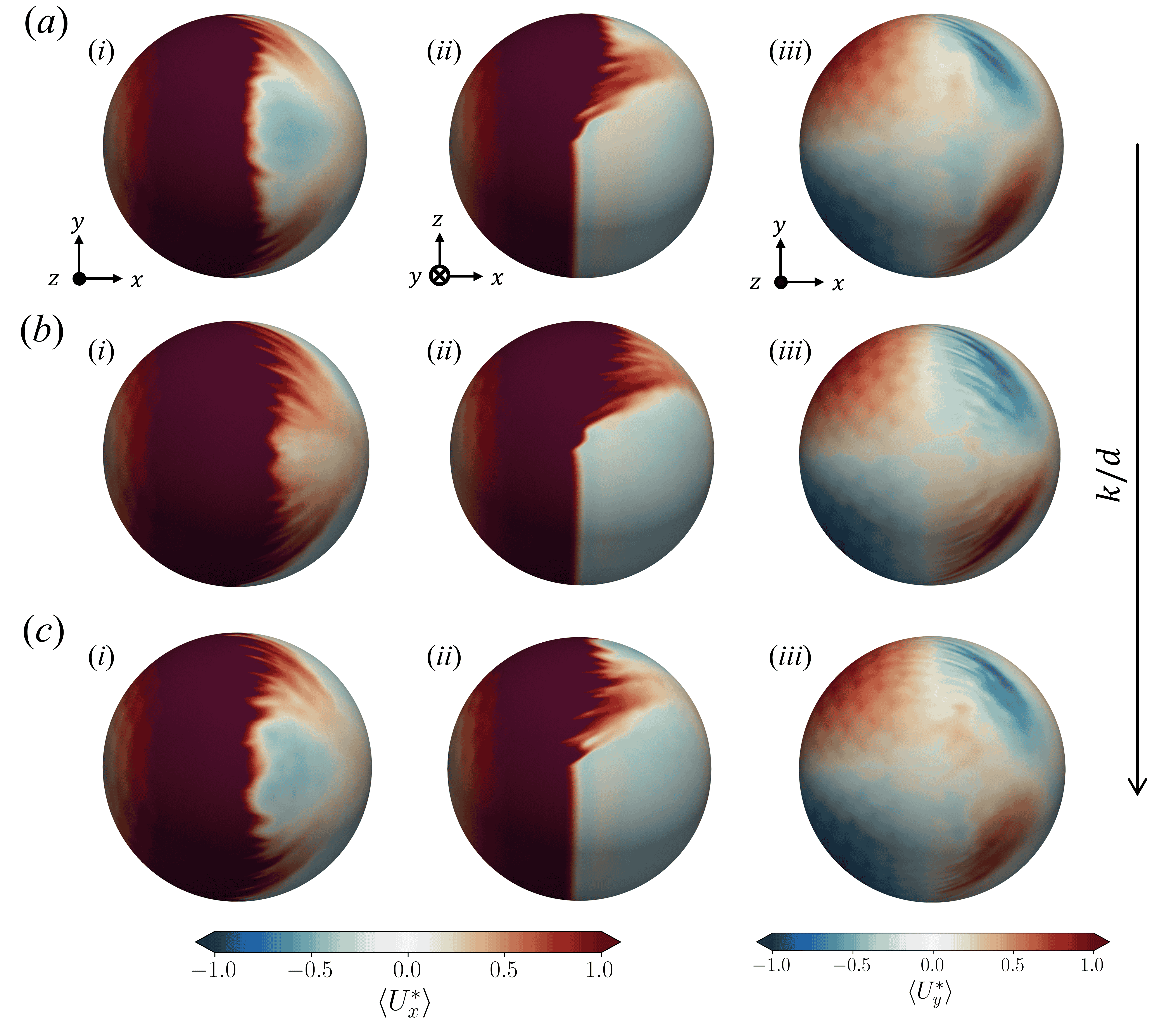}
    \caption{Time-averaged near-wall velocity fields for the smooth and asymmetric dimpled spheres at $Re=100{,}000$. Panels show the normalized streamwise velocity, $\langle U_x^* \rangle$, from the (i) top and (ii) side views, and the normalized spanwise velocity, $\langle U_y^* \rangle$, from the (iii) top view. Results are shown for (a) the smooth sphere, (b) the asymmetric dimpled sphere with $k/d=0.004$, and (c) the asymmetric dimpled sphere with $k/d=0.006$. The velocity is plotted near the sphere surface at $r/d=0.51$.}

    \label{fig:uxuywall}
\end{figure}

The sidewash mentioned above and the local separation region can be seen more clearly in the contours of the time-averaged streamwise velocity $\langle U_x^*\rangle$ and spanwise velocity $\langle U_y^*\rangle$ plotted right above the surface at $r/d=0.51$ in Fig.~\ref{fig:uxuywall} for the three asymmetrically dimpled cases. For $k/d=0.004$ and $k/d=0.008$, $\langle U_x^*\rangle$ contours clearly show local recirculation at the top, which is encapsulated by the strongly delayed separation region with a positive streamwise velocity (Fig.~\ref{fig:uxuywall}a(i) and ~\ref{fig:uxuywall}c(i)). For both of these cases, positive values of $\langle U_x^*\rangle$ further downstream indicate partial flow attachment. The sidewash due to the pressure difference between the smooth and the dimpled side akin to tip vortices in finite wings can also be seen in Fig.~\ref{fig:uxuywall}a(ii) and ~\ref{fig:uxuywall}c(ii) which combined with the surface curvature leads to spanwise converging flow as seen in Fig.~\ref{fig:uxuywall}a(iii) and ~\ref{fig:uxuywall}c(iii) for $k/d=0.004$ and $k/d=0.008$ respectively. In contrast, for the case with maximum lift i.e. $k/d=0.006$, this local recirculation region shrinks significantly as sidewash becomes stronger (Fig.~\ref{fig:uxuywall}b(i-iii)).


To further investigate the flow separation on the dimpled hemisphere in the wall-normal direction, the tangential velocity profiles normalized by the freestream velocity $\langle u_\phi\rangle/U_\infty$ are plotted along the radial direction $r/d$ in Fig.~\ref{fig:cpvelprofile} at $\theta=90^\circ$. A velocity overshoot, $\langle u_\phi\rangle/U_\infty>1$, is observed near the wall in all cases with dimples, which is marginal for the smooth sphere. This suggests that the velocity overshoot is likely due to dimple-induced jetting and turbulent mixing.  At $\phi=90^\circ$ where the flow is already separated in the smooth sphere, the dimpled cases show attached flow at this $\theta$ (Fig.~\ref{fig:cpvelprofile}e,f). The velocity profiles for $k/d=0.004$ and $k/d=0.008$ are similar: reversed flow appears at $\phi=110^\circ$ and persists through $\phi=120^\circ$, with a wall-normal depth of approximately $0.03d$ at both cases at $\phi=120^\circ$. The mean lift coefficient for these two cases is also similar (Table~\ref{tab:force_comp1}), which is expected from the similarity of the velocity profiles and the separation region for these two cases. In contrast, for $k/d=0.006$, the reversed flow at $\phi=110^\circ$ is restricted to a wall-normal depth less than $0.01d$, and the profile at $\phi=120^\circ$ remains positive throughout, indicating that separation does not reach this azimuthal location. Together with Fig.~\ref{fig:uxuywall}, this indicates a much smaller separation region for $k/d=0.006$ in both the azimuthal and wall-normal directions.

One consistent feature in both the experiments~\cite{sudarsana2024} and the present LES is that $\langle C_D\rangle$ remains weakly dependent on the dimple depth $k/d$ (Table~\ref{tab:force_comp1}), despite substantial differences in flow structure and significant changes in the mean lift coefficient. The reason behind this trend was not answered in the experimental study due to limited diagnostics. To explain this, the drag is decomposed into pressure and viscous contributions, as well as front- and rear-hemisphere components. The net hydrodynamic force on the sphere can be written as:
\begin{equation}
\label{force}
    \mathbf{F} = \int_{S_b} (-p\,\mathbf{n} + \boldsymbol{\tau}\cdot \mathbf{n} ) dS,
\end{equation}
where $S_b$ is the sphere surface and $\mathbf{n}$ is the outward unit normal. In the freestream direction $\hat{\mathbf{e}}_x$, drag can be expressed as $D=-\mathbf{F}\cdot \hat{\mathbf{e}}_x$, therefore the drag coefficient:
\begin{equation}
\label{CD}
    C_D = \underbrace{\frac{2}{\rho U_\infty^2 A}\int_{S_b} p\,(\mathbf{n}\cdot \hat{\mathbf{e}}_x) \ dS}_{C_{D_p}} - \underbrace{\frac{2}{\rho U_\infty^2A}\int_{S_b} (\boldsymbol{\tau}\cdot\mathbf{n}) \cdot \hat{\mathbf{e}}_x \ dS.}_{C_{D_f}}
\end{equation}

As shown in Table~\ref{tab:force_comp1}, approximately 98\% of the total drag in all cases is due to pressure drag, $C_{D_p}$, and this contribution changes only slightly by adding dimples. This is expected since a sphere is a bluff body, and the pressure drag dominates the total drag for bluff bodies. The pressure drag is further decomposed into front and rear contributions as:

\begin{equation}
\label{CDp}
    C_{D_p} = \underbrace{\frac{2}{\rho U_\infty^2 A}\int_{S_{b, F}} p\,(\mathbf{n}\cdot \hat{\mathbf{e}}_x) \ dS}_{C_{D_{p,F}}} + \underbrace{\frac{2}{\rho U_\infty^2 A}\int_{S_{b, R}} p\,(\mathbf{n}\cdot \hat{\mathbf{e}}_x) \ dS}_{C_{D_{p, R}}},
\end{equation}
where $C_{D_{p,F}}$ and $C_{D_{p,R}}$ denote the pressure drag on the front and rear hemispheres, respectively and $S_{b,F}$ corresponds to surface area of the front hemisphere ($\phi=0^\circ-90^\circ$ and $270^\circ-360^\circ$), while $S_{b,R}$ corresponds to surface area of the rear hemisphere ($\phi=90^\circ-270^\circ$). Table~\ref{tab:force_decom} shows that although the total pressure drag is nearly identical for all the cases, its spatial distribution differs; the smooth sphere has a larger $C_{D_{p,F}}$, whereas the dimpled sphere has a larger $C_{D_{p,R}}$ which changes slightly with the dimple depth. Together with Fig.~\ref{fig:cpsmodim}, this indicates that a sphere with asymmetric dimples primarily redistributes the pressure loading between the front and rear hemispheres rather than reducing the total drag.

\begin{table*}[htbp]
\caption{\label{tab:force_decom}Decomposition of the pressure-drag coefficient into front- and rear-hemisphere contributions for smooth and asymmetrically dimpled spheres at $Re$=100,000. Although asymmetric roughness redistributes the pressure-drag contribution between the front and rear hemispheres, the total pressure-drag coefficient remains nearly unchanged. Here, subscripts F and R denote the front and rear hemispheres, respectively.}
\begin{tabular}{lccccc}
\hline
      ~~~~~~~~ & $\langle C_{D_{p}}\rangle$ & $\langle C_{D_{p,F}}\rangle$ & $\langle C_{D_{p,R}} \rangle$ & $\langle C_{D_{p,F}}\rangle/\langle C_{D_{p}}\rangle$ & $\langle C_{D_{p,R}} \rangle/\langle C_{D_{p}}\rangle$ \\
      \hline
       Smooth & 0.4954 & 0.0929 & 0.4025 & 18.76\% & 81.24\% \\
       Asymmetric dimpled $k/d=0.004$ & 0.5095 & 0.022 & 0.4875 & 4.32\% & 95.68\% \\
       Asymmetric dimpled $k/d=0.006$ & 0.4932 & 0.0073 & 0.486 & 1.48\% & 98.54\% \\
       Asymmetric dimpled $k/d=0.008$ & 0.4959 & 0.0251 & 0.470 & 5.06\% & 94.77\% \\
       \hline
\end{tabular}
\end{table*}

Equation~\ref{CD} also shows that drag depends only on the streamwise projection of the surface force. Therefore, a significant drag reduction would require a substantial change in the surface pressure distribution integrated with the $\mathbf{n}\cdot\hat{\mathbf{e}}_x$ weighting over the whole surface. As shown in Table \ref{tab:force_decom}, asymmetric dimple perturbation only redistributes the front and rear pressure loading. Moreover, Fig.~\ref{fig:3dsepsmodim} shows that the delay in separation occurs only on the dimpled side, varying at $\phi_s\sim80^\circ-120^\circ$, which are still collectively smaller than the $\phi_s\sim110^\circ$ reported for a fully dimpled golf-ball-like sphere~\cite{Choi2006}. Thus, the net change in the streamwise projection of the pressure distribution remains limited. This is further elaborated by decomposing the surface pressure field into axisymmetric and non-axisymmetric terms using a Fourier decomposition. For this decomposition, a streamwise spherical coordinate system $(\theta, \phi)$ is introduced, where $\theta$ is the streamwise polar angles measured from the upstream stagnation point $-\hat{\mathbf{e}}_x$ with a range $[0,\pi]$, and $\phi$ is the azimuthal angles around the $x$-axis, measured from  $-\hat{\mathbf{e}}_y$ with a range of $[0,2\pi)$. Therefore, the pressure decomposition can be expressed as
\begin{align}
    p(\theta, \phi) = \langle{p}(\theta)\rangle + p'(\theta,\phi),
\end{align}
where $\langle{p}(\theta)\rangle$ represents the azimuthal average or the $m=0$ Fourier mode, defined as $\langle{p}(\theta)\rangle=\frac{1}{2\pi}\int_0^{2\pi} p(\theta,\phi) d\phi$ and $p'(\theta,\phi)$ is the sum of all non-axisymmetric Fourier modes $m=1,2,3,\cdots$ which carries all pressure variation across $\phi$ for each $\theta$. Physically, $\langle{p}(\theta)\rangle$ is the front-to-back axisymmetric pressure distribution, while $p'(\theta,\phi)$ is the transverse asymmetric pressure variation. Following the pressure drag definition in Eq. \ref{CD}, with $(\mathbf{n}\cdot \hat{\mathbf{e}}_x)= \cos{\theta}$ and $dS=R^2\sin{\theta} \, d\theta \, d\phi$, the drag decomposition yields

\begin{align}
\label{Fxdecom}
    F_{x,p} =  \underbrace{R^2 \int_0^\pi \int_0^{2\pi} \langle{p}(\theta)\rangle \cos{\theta} \sin{\theta} \, d\phi \, d\theta}_{F^{(0)}_{x,p}} + \underbrace{R^2 \int_0^\pi \int_0^{2\pi} p'(\theta, \phi) \cos{\theta} \sin{\theta} \, d\phi \, d\theta}_{F'_{x,p}} 
\end{align}
The decomposition of pressure field yields to two terms of drag, the axisymmetric term $F^{(0)}_{x,p}$ and the asymmetric term $F'_{x,p}$. By definition, the azimuthal integral of $p'(\theta,\phi)$ is $\int_0^{2\pi} p'(\theta,\phi)d\phi=0$ since $p'(\theta,\phi)$ has zero azimuthal mean around each $\theta$, therefore only the first term of Eq. \ref{Fxdecom} governs the pressure drag, 
\begin{align}
    \label{Fx}
     F_{x,p} =  2\pi R^2  \int_0^\pi  \langle{p}(\theta)\rangle \cos{\theta} \sin{\theta} \,  d\theta
\end{align}
This implies that the pressure drag is solely defined by the azimuthally averaged pressure distribution $\langle{p}(\theta)\rangle$. For the pressure term of lift, the pressure is integrated with the weighting of $(\mathbf{n}\cdot \hat{\mathbf{e}}_z)=\sin{\theta} \sin{\phi}$, therefore
\begin{align}
    F_{z,p} = \underbrace{-R^2 \int_0^\pi \int_0^{2\pi} \langle{p}(\theta)\rangle \sin^2{\theta} \sin{\phi} \, d\phi \, d\theta}_{F^{(0)}_{z,p}} \ \ \underbrace{ - \, R^2 \int_0^\pi \int_0^{2\pi} p'(\theta, \phi) \sin^2{\theta} \sin{\phi} \, d\phi \, d\theta}_{F'_{z,p}}
\end{align}
Similarly as drag in Eq. \ref{Fxdecom}, lift is decomposed by its pressure axisymmetric term $F^{(0)}_{z,p}$ and the asymmetric term $F'_{z,p}$. The first term vanishes due to $\int_0^{2\pi}\sin{\phi} \, d\phi=0$, therefore the pressure term of lift is only governed by the non-axisymmetric part of the decomposed pressure,
\begin{align}
    \label{Fz}
    F_{z,p} = -R^2 \int_0^\pi \int_0^{2\pi} p'(\theta, \phi) \sin^2{\theta} \sin{\phi} \, d\phi \, d\theta
\end{align}

As intuitively understood, the decomposition of pressure yields Equations \ref{Fx} and \ref{Fz} which mathematically demonstrate that for an ideal spherical object, the drag is governed only by the axisymmetric surface pressure field, while for lift, the asymmetric pressure variation dominates. Table \ref{tab:force_decom_fourier} tabulates the decomposition of the drag coefficient, consisting of the axisymmetric $\langle C^{(0)}_{D_{p}}\rangle$ and the asymmetric terms $\langle C'_{D_{p}}\rangle$, and the lift coefficient $\langle C^{(0)}_{L_{p}}\rangle$ and $\langle C'_{L_{p}}\rangle$. The axisymmetric term $\langle p(\theta) \rangle$ requires azimuthal average and integral approximation which is evaluated by grouping the cells into a narrow bins $\theta$, where for each bin $\theta_i$, an area-weighted average pressure is
\begin{align}
    \langle p(\theta_i) \rangle = \frac{\sum_{f\in \theta_i} p_f \Delta S_f}{\sum_{f\in \theta_i} \Delta S_f},
\end{align}
where $f$ denotes surface faces for one bin $\theta$ and $\Delta S_f$ is the area of each surface face. From this, the asymmetric term can be obtained from $p'(\theta,\phi)=p(\theta,\phi) - \langle p(\theta_i) \rangle$. As expected, the asymmetric term of the drag is nearly zero for all cases $\langle C^{(0)}_{D_{p}}\rangle\approx0$ and the axisymmetric term is similar to the total pressure drag $\langle C^{(0)}_{D_{p}}\rangle\approx\langle C_{D_{p}}\rangle$. This clearly demonstrates that the asymmetric term contributes only a minimal amount, if any, to the total drag. For the pressure term of lift, all cases produce $\langle C^{(0)}_{L_{p}}\rangle \approx0$, while for all $k/d$ cases, the asymmetric term is close to the total lift $\langle C^{(0)}_{L_{p}}\rangle\approx\langle C_{L_{p}}\rangle$. Although the asymmetric term contributes only minimally to the total drag, it generates significant lift. Physically, this implies that asymmetric dimple perturbation does not alter the streamwise-projected pressure distribution, even though it produces a significant lift or transverse projection of the pressure distribution.

\begin{table*}[htbp]
\caption{\label{tab:force_decom_fourier} Force decomposition (pressure) on the body via Fourier decomposition of the surface pressure field on a smooth and dimpled sphere at $Re=100{,}000$. The shaded color differentiates values that are $\approx0$ (red) and non-trivial (green).}
\centering
\begin{tabular}{lcccc}
\hline
 & $\langle C^{(0)}_{D_{p}}\rangle$ 
 & $\langle C'_{D_{p}}\rangle$ 
 & $\langle C^{(0)}_{L_{p}}\rangle$ 
 & $\langle C'_{L_{p}}\rangle$ \\
\hline

Smooth 
& \cellcolor{largegreen}0.495241 
& \cellcolor{smallred}$7.5\times 10^{-5}$ 
& \cellcolor{smallred}$8.27\times10^{-11}$ 
& \cellcolor{smallred}$-0.918\times10^{-2}$ \\ 

Asymmetric dimpled $k/d=0.004$ 
& \cellcolor{largegreen}0.508 
& \cellcolor{smallred}$0.127\times10^{-2}$ 
& \cellcolor{smallred}$-0.745\times10^{-3}$ 
& \cellcolor{largegreen}0.319942 \\

Asymmetric dimpled $k/d=0.006$ 
& \cellcolor{largegreen}0.4896 
& \cellcolor{smallred}$0.327\times10^{-2}$ 
& \cellcolor{smallred}$-0.911\times10^{-3}$ 
& \cellcolor{largegreen}0.359419 \\

Asymmetric dimpled $k/d=0.008$ 
& \cellcolor{largegreen}0.4907 
& \cellcolor{smallred}$0.481\times10^{-2}$ 
& \cellcolor{smallred}$-0.125\times10^{-2}$ 
& \cellcolor{largegreen}0.316656 \\

\hline
\end{tabular}
\end{table*}


\subsection{\label{sec3sub2}Near-wake structure and turbulence}
 
\begin{figure}[htbp]
    \centering
    \includegraphics[width=1\linewidth]{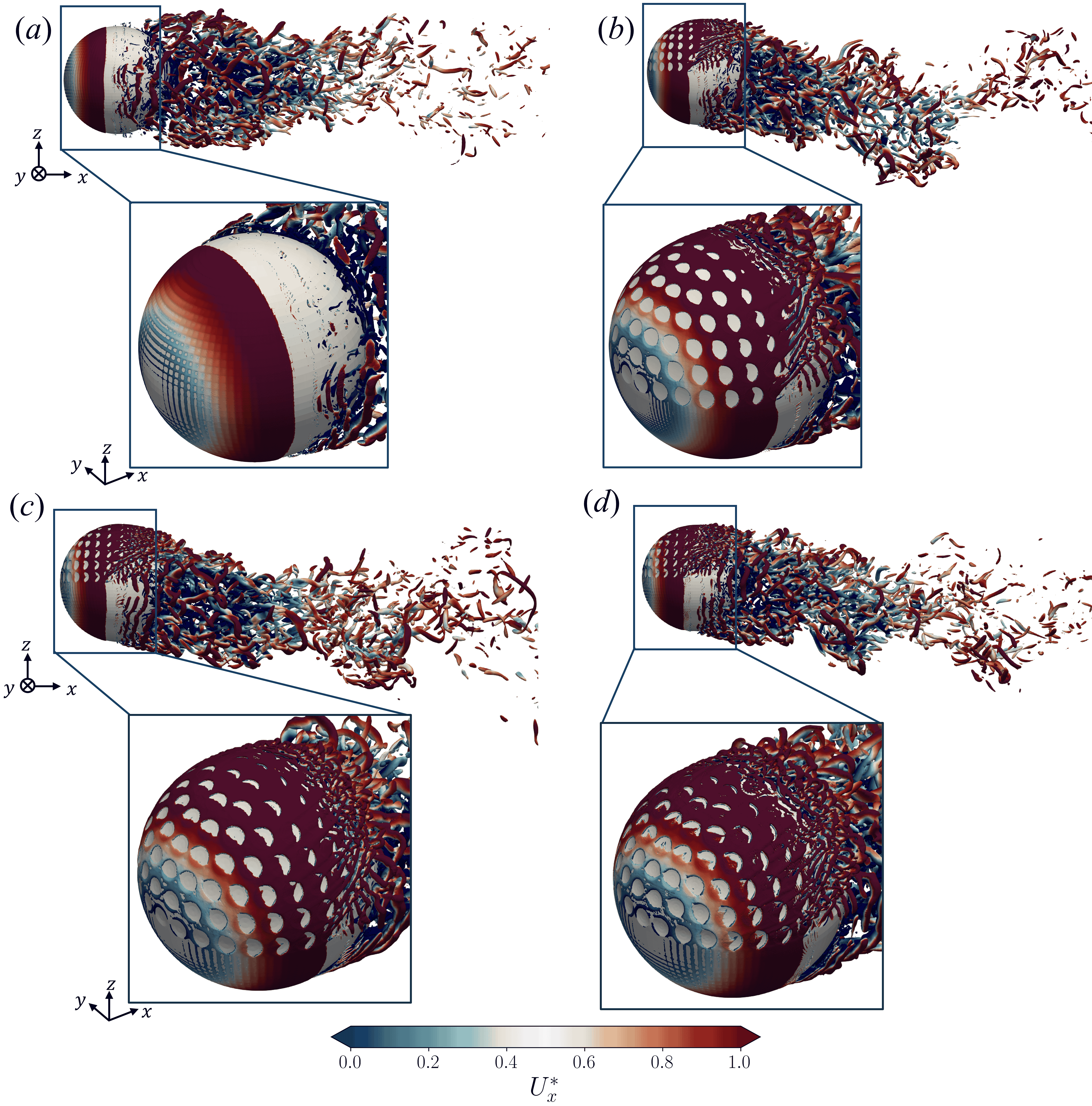}
    \caption{Instantaneous vortical structures identified using iso-surfaces of the $Q$-criterion at $Q=100$ for (a) the smooth sphere and asymmetric dimpled spheres with (b) $k/d=0.004$, (c) $k/d=0.006$, and (d) $k/d=0.008$. The iso-surfaces are colored by the instantaneous normalized streamwise velocity, $U_x^*$. Each panel includes a full isometric view of the wake and a close-up view highlighting the near-surface flow structures around the dimpled region.}

    \label{fig:qcrit}
\end{figure}

The three-dimensional separation behavior (varying separation locations) due to the modified surface pressure coefficient $\langle C_P\rangle$ distribution, as discussed in the previous section, suggests complex near-wake structure dynamics. Figure \ref{fig:qcrit} shows the vortical structures identified using the instantaneous iso-surfaces of $Q$, where $Q=\frac{1}{2}(\boldsymbol{\omega}_{ij}\boldsymbol{\omega}_{ij} - \mathbf{S}_{ij}\mathbf{S}_{ij})$ with $\boldsymbol{\omega}=\frac{1}{2}(\nabla \mathbf{u} - (\nabla \mathbf{u})^\top)$ and $\mathbf{S}=\frac{1}{2}(\nabla \mathbf{u} + (\nabla \mathbf{u})^\top)$, is the second invariant of the velocity gradient tensor that measures the excess of local rotation rate compared to the strain rate \cite{Hunt1988, Jeong_Hussain_1995}. Figure~\ref{fig:qcrit}a shows that the smooth sphere exhibits symmetric vortical structures about the $x$ axis, which is expected since no lift is generated in this case. Moreover, high vortical structure regions are concentrated in the boundary layer and in the separated shear layer further downstream, with smaller structures more abundant in the shear layer (3D close-up view of Fig.~\ref{fig:qcrit}a), indicating that the transition to turbulence occurs at far downstream in the shear layer. In contrast, the sphere with asymmetric dimples exhibits wake deflection (Fig.~\ref{fig:qcrit}b-d). The close-up view in Fig.~\ref{fig:qcrit}b-d shows small vortical structures typical of a turbulent separated shear layer, with their size and inclination varying with $\theta$ on the dimpled hemisphere. This qualitatively shows that the formation of small vortical structures at different $\theta$ and $\phi$ angles is strongly dependent on the separation location on this hemisphere. Further quantification of the wake deflection is obtained from the time-averaged streamwise velocity $\langle U_x \rangle$ at the midplane. As expected, the smooth sphere case for which no mean lift was generated, no wake deflection is observed, while the asymmetric dimpled spheres $k/d=0.004$ and $0.008$ exhibit $\sim11.3^\circ$ wake deflection about the $+x$-axis. The case with the maximum lift, i.e., $k/d=0.006$, shows the maximum wake deflection with $\sim13.13^\circ$ (see Appendix \ref{appendix2}).

\begin{figure}[htbp]
    \centering
    \includegraphics[width=0.8\linewidth]{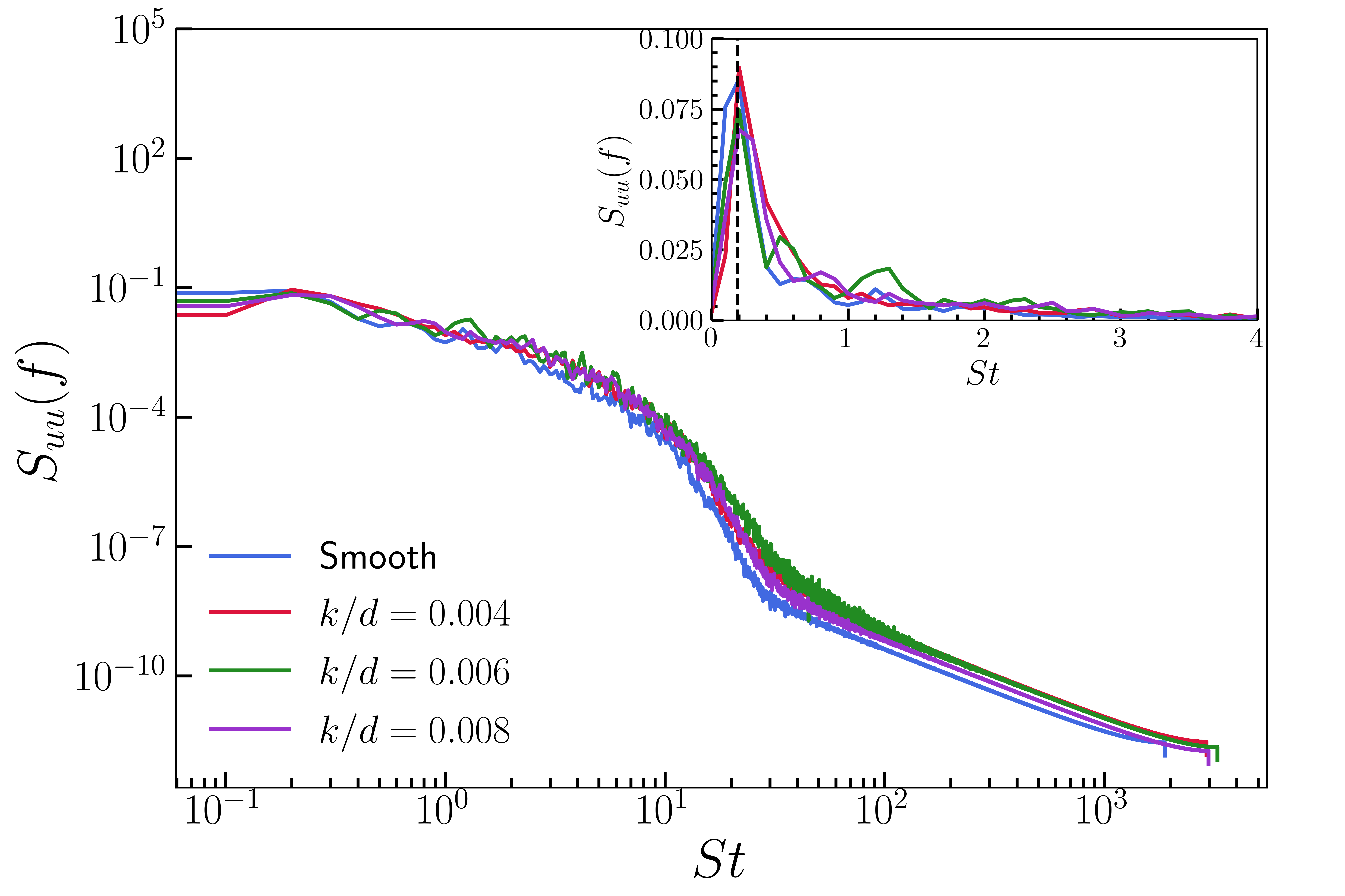}
    \caption{Power Spectral Density (PSD) $S_{uu}(f)$ for smooth sphere and  asymmetric dimpled spheres with different $k/d$. The probe is located at $(x,y,z)$ = $(2d,0d,-0.5d)$.}
    \label{fig:psd}
\end{figure}

Figure \ref{fig:psd} shows the power spectral density (PSD) $S_{uu}(f)$ of the instantaneous streamwise velocity $u_x(t)$ for the smooth sphere and the three asymmetrically dimpled cases, where $S_{uu}(f)=|\hat u(f)|^2/\tau$ where $\hat u(f)$ is the Fourier transform of a point data signal $u(t)$ located at $(x,y,z)=(2d, 0d, -0.5d)$. Welch's method is used to estimate the signal's PSD with a 50\% overlap and a Hamming window. The smooth and dimpled spheres show no significant difference in dominant shedding frequency. Both show a peak in $S_{uu}(f)$ at $St\approx0.19$, which agrees well with previous studies of a smooth sphere \cite{Achenbach1974a, Achenbach1974b, vilumbrales2025}. This indicates that while asymmetric dimples induce a wake deflection toward $-z$, opposite to the lift, they do not significantly affect the shedding frequency. Achenbach \cite{Achenbach1974a} also reported an insignificant change in $St$ between smooth and fully non-uniformly roughened (protrusion) spheres across different parameter $k/d$. This suggests that while surface roughness (or dimpling) in half- or full-configuration can alter boundary-layer behavior on a sphere, the effect on vortex-shedding characteristics remains limited.

\begin{figure}[htbp]
    \centering
    \includegraphics[width=1\linewidth]{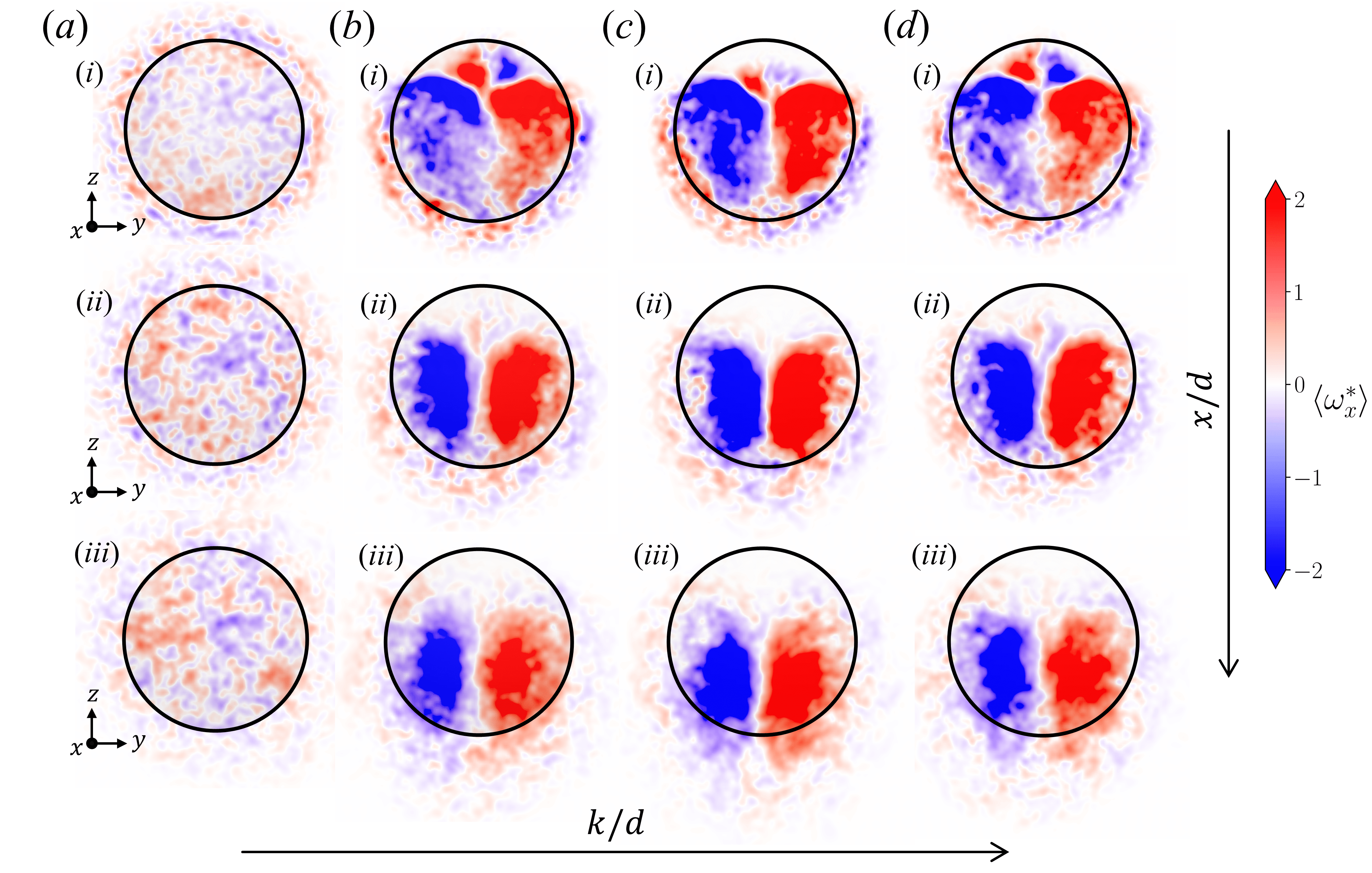}
    \caption{Time-averaged normalized streamwise vorticity, $\langle \omega_x^* \rangle$, in cross-sectional wake planes for (a) the smooth sphere and asymmetric dimpled spheres with (b) $k/d=0.004$, (c) $k/d=0.006$, and (d) $k/d=0.008$. Results are shown at three downstream locations: (i) $x/d=0.6$, (ii) $x/d=1.0$, and (iii) $x/d=1.5$.}
    \label{fig:vortx_cross}
\end{figure}

To characterize the vorticity in the wake, the time-averaged normalized streamwise vorticity $\langle \omega_x^*\rangle$ is plotted for the smooth and dimpled spheres in Fig.~\ref{fig:vortx_cross} at three different streamwise locations $x/d=0.6$, $x/d=1$, and $x/d=1.5$. The asymmetric dimpled spheres show the formation of a counter-rotating vortex pair (Fig.~\ref{fig:vortx_cross}b-d), which is not present in the smooth sphere (Fig.~\ref{fig:vortx_cross}a). For $k/d=0.004$ and $k/d=0.008$, a double counter-rotating vortex pair is formed at $x/d=0.6$, with a smaller secondary vortex pair on top of the primary vortex pair. The secondary vortex pair appears weaker for $k/d=0.006$ at $x/d=0.6$. This secondary vortex pair dissipates and merges into the primary vortex pair by $x/d=1$. For all asymmetrically dimpled cases, the vortex pair moves downward in the opposite direction of lift ($-z$) as it is convected downstream, a clear indication of wake deflection. Initial signs of these counter-rotating vortex pairs are also evident in the skin friction lines in Fig.~\ref{fig:surfaceplot} and Fig.~\ref{fig:cpvelprofile}. The formation of a counter-rotating vortex pair was also observed in a previous study of a cricket ball \cite{Parekh2024} and a rotating sphere \cite{Milner2025}.

\begin{figure}[htbp]
    \centering
    \includegraphics[width=1\linewidth]{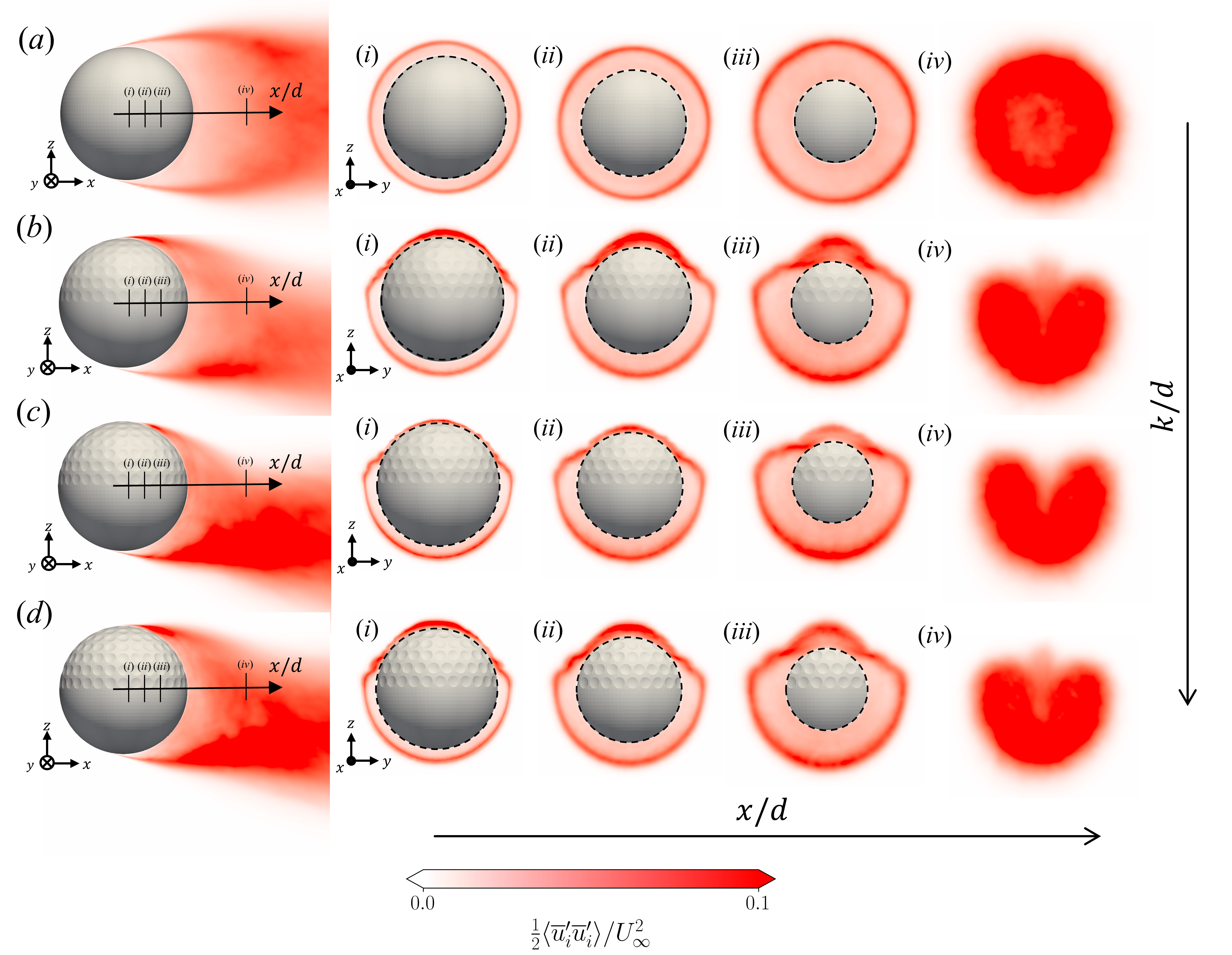}
    \caption{Normalized resolved turbulent kinetic energy $\frac{1}{2}\langle \overline{u}'_i \overline{u}'_i\rangle/U_{\infty}^2$ for (a) Smooth, (b) Asymmetric dimpled $k/d=0.004$, (c) Asymmetric dimpled $k/d=0.006$, and (d) Asymmetric dimpled $k/d=0.008$, in different cross-plane near-wake locations (i) $x/d=0.2$, (ii) $x/d=0.3$, (iii) $x/d=0.4$, and (iv) $x/d=1$ along with the intersection of the sphere for (i)-(iii). The leftmost side of the figure shows the mid-plane view (streamwise normal) with the corresponding locations of (i)-(iv). }
    \label{fig:tke_cross}
\end{figure}

Due to the difference in the roughness on the two sides of the sphere and as indicated by the three-dimensional wake structure described above, the boundary layer on each hemisphere is expected to transition through different pathways on the smooth and dimpled sides. Turbulent kinetic energy (TKE), $k=\frac{1}{2}\langle \overline{u}'_i \overline{u}'i\rangle/U_\infty^2$, provides a direct measure of this asymmetry, as shown in Fig.~\ref{fig:tke_cross} for successive cross-planes. On the smooth sphere, high TKE appears in the separated shear layer, downstream of the separation point (Fig.~\ref{fig:tke_cross}a). This is consistent with the well-known behavior in which a laminar boundary layer separates and subsequently transitions in the free shear layer \cite{Achenbach1972, Deshpande2017}. In contrast, on the dimpled sphere, high TKE is concentrated near the surface on the dimpled side and upstream of the separation location, whereas the smooth hemisphere shows the same shear-layer transition as observed in the smooth sphere (Fig.~\ref{fig:tke_cross}b-d(i)). This near-wall concentration of TKE persists in the downstream cross-planes (Fig.~\ref{fig:tke_cross}b-d(ii,iii)) and is localized around $\theta\sim60^\circ$--$120^\circ$, coinciding with the separation region identified in Fig.~\ref{fig:3dsepsmodim}. The appearance of high TKE close to the sphere surface rather than in the free shear layer indicates that the dimpled-side boundary layer undergoes transition to turbulence, which delays separation. This observation aligns well with several studies that have shown that dimples trigger the transition to turbulence in flow past a sphere \cite{Choi2006, Beratlis2019}.   

\begin{figure}[htbp]
    \centering
    \includegraphics[width=0.8\linewidth]{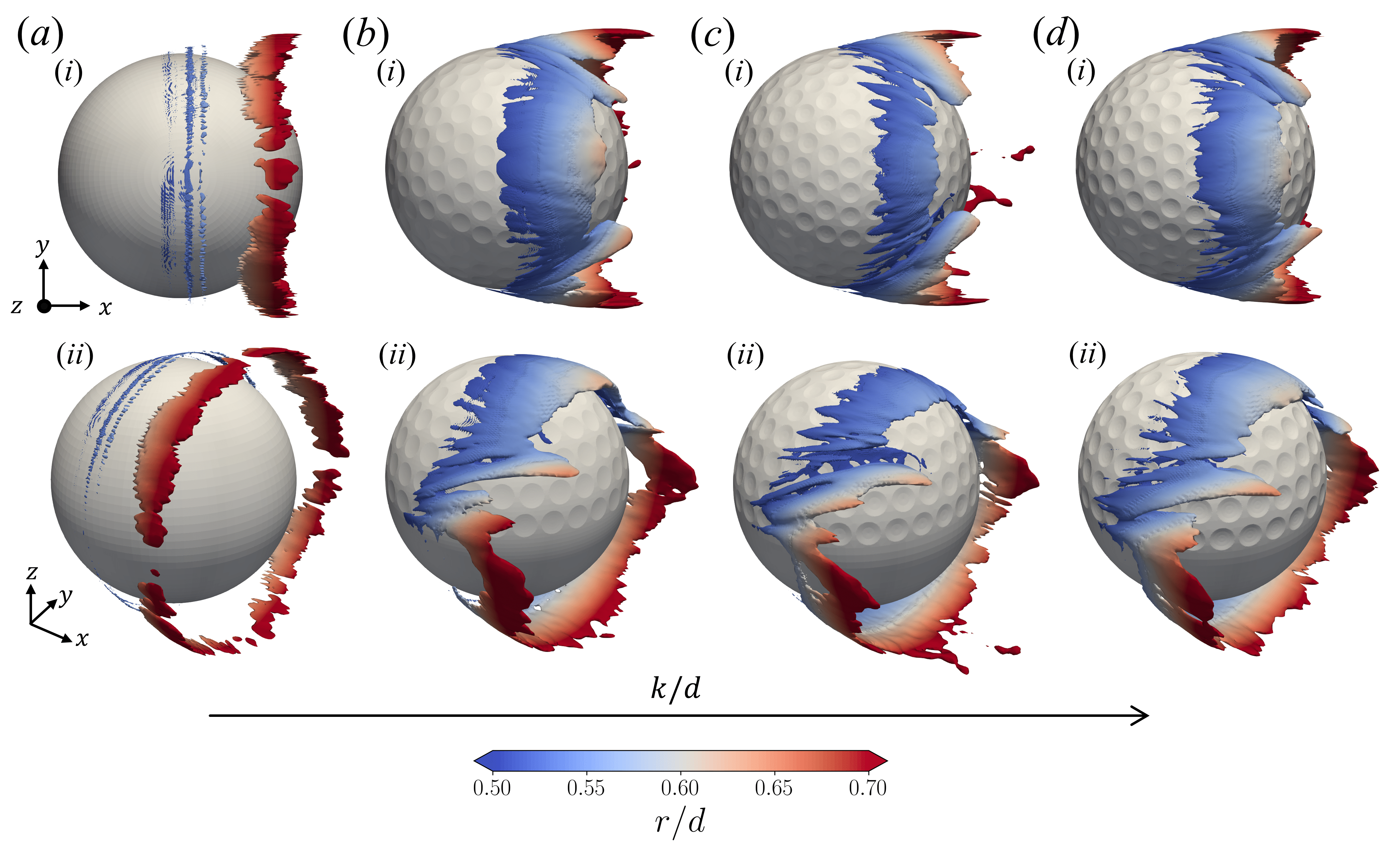}
    \caption{Regions of high turbulent kinetic energy production identified via $\mathcal{P}_kD/U_\infty^3\approx0.5$ colored by the radial distance from the sphere center for (a) Smooth, (b) Asymmetric dimpled $k/d=0.004$, (c) Asymmetric dimpled $k/d=0.006$, and (d) Asymmetric dimpled $k/d=0.008$. (i) corresponds to the top view and (ii) corresponds to the 3D view.} 
    \label{fig:ptke_3d}
\end{figure}

TKE production further characterizes the transition on each hemisphere. The TKE budget equation for an incompressible flow can be written as:
\begin{equation}
\frac{Dk}{Dt} = \underbrace{-\langle u_i'u_j'\rangle \frac{\partial \langle u_i \rangle}{\partial x_j}}_{\mathcal{P}_k} + \mathcal{T}_k - \varepsilon,
\end{equation}
where $\mathcal{P}_k$ is the production of TKE by the mean strain, $\mathcal{T}_k$ groups the turbulent transport, pressure diffusion, and viscous diffusion, and $\varepsilon$ is the viscous dissipation rate \cite{Pope2000}. For the current analysis, only the production term, $\mathcal{P}_k$, is relevant. The non-dimensionalized form can be expressed as $\mathcal{P}_k^*=\mathcal{P}_kD/U_\infty^3$. Figure~\ref{fig:ptke_3d} compares the iso-surface of $\mathcal{P}_k^* \approx 0.5$  colored by the radial distance from the sphere center for all the cases. On the smooth sphere, the turbulence production is confined to the free shear layer downstream of separation and away from the surface of the sphere (Fig.~\ref{fig:ptke_3d}a), where the detached laminar layer rolls up and transitions to turbulence. The smooth hemisphere of the asymmetrically dimpled sphere shows similar behavior. On the dimpled hemisphere, turbulence production occurs much closer to the surface, however, the intense TKE production occurs only after separation, as indicated by the colored contours of $\mathcal{P}_k^*$ shifting downstream (Fig.~\ref{fig:ptke_3d}). The radial distance contour shows that turbulence production in $k/d=0.006$ is further delayed and concentrated in the near-wall region at $\theta\sim50^\circ-150^\circ$ (Fig.~\ref{fig:ptke_3d}c(i)), compared to other $k/d$ cases (Fig.~\ref{fig:ptke_3d}b(i) and Fig.~\ref{fig:ptke_3d}d(i)). These results indicate that two distinct transition pathways coexist on the asymmetrically dimpled sphere: the smooth-side boundary layer separates while still laminar and transitions in the free shear layer, whereas the dimpled-side boundary layer becomes turbulent before it separates. The separated shear layers on the two hemispheres consequently begin with very different turbulent states at their respective points of detachment.

\subsection{\label{sec3sub4}Mechanism of enhanced wake deflection and lift generation}

\begin{figure}[htbp]
    \centering
    \includegraphics[width=0.8\linewidth]{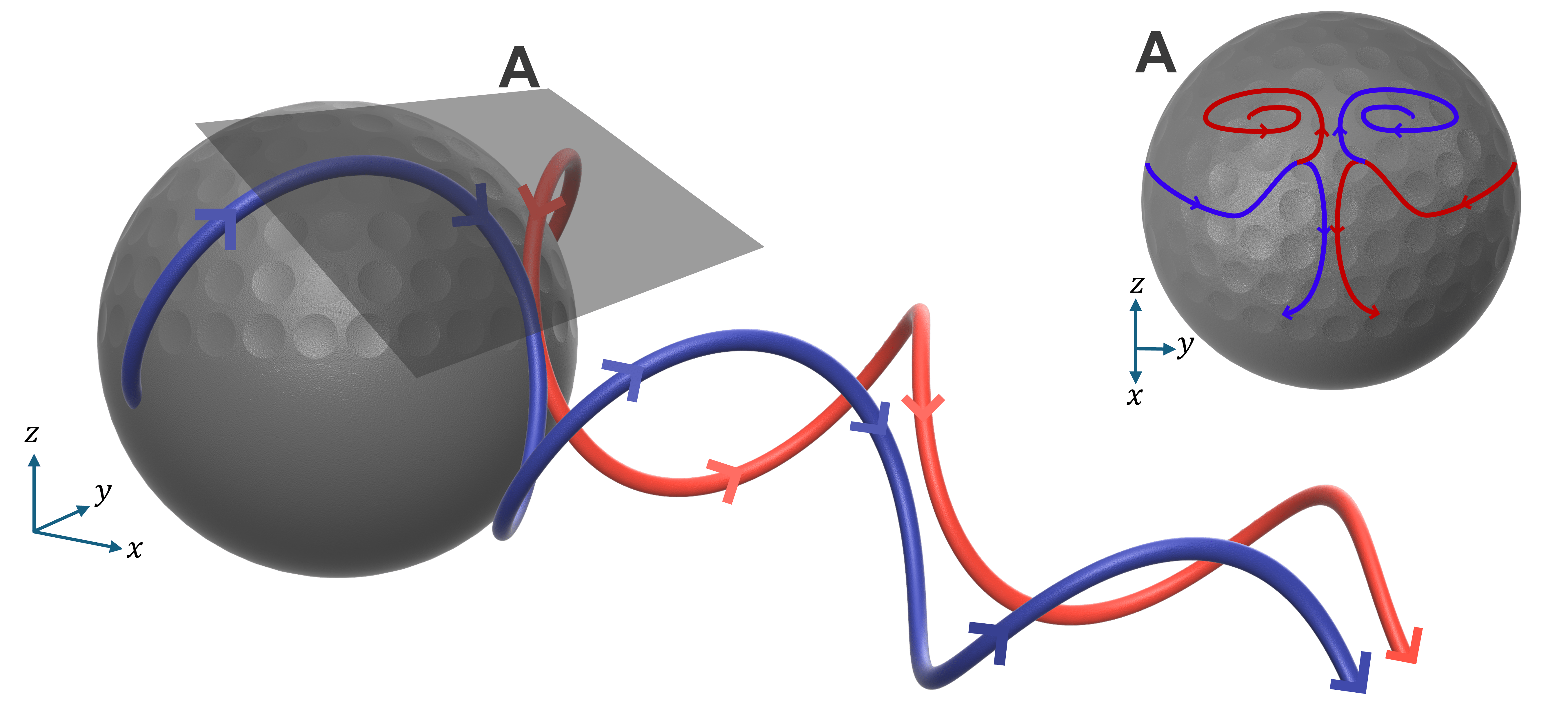}
    \caption{Schematic showing the generation of counter-rotating vortex pair and a secondary vortex-pair via sidewash over an asymmetrically dimpled sphere at $Re=100{,}000$.}
    \label{fig:illustration}
\end{figure}

As established in Sec.~\ref{sec3sub1} and Sec.~\ref{sec3sub2}, asymmetric dimples trigger boundary-layer transition and delay separation non-uniformly on the dimpled hemisphere to $\phi_s\sim100^\circ$--$130^\circ$, while the smooth hemisphere separates at $\phi_s\sim80^\circ$, producing the surface pressure asymmetry visible in Figs.~\ref{fig:cpsmodim} and~\ref{fig:uxuywall}. This pressure asymmetry drives a cross-stream sidewash from the smooth toward the dimpled hemisphere, as visible in the $\langle U_y \rangle$ contours of Fig.~\ref{fig:uxuywall}, which rolls up downstream into the counter-rotating streamwise vortex pair observed in Fig.~\ref{fig:vortx_cross}. The mechanism is analogous to the formation of tip vortices on a finite wing, where the pressure difference between the pressure and suction sides drives spanwise flow around the wing tip \cite{Parekh2024, Milner2025}. The cores of this vortex pair produce a downward induced velocity via the Biot--Savart law. The TKE cross-plane at $x/d=1$ (Fig.~\ref{fig:tke_cross}b-d(iv)) confirms this vortex-driven reorganization: the high-TKE shear layers originating on the dimpled side are wrapped by the counter-rotating vortex pair into a heart-shaped distribution, demonstrating that the vortex pair actively reshapes the near-wake. The wake deflection therefore reflects the combined action of the spatially non-uniform, three-dimensional separation topology, the pressure asymmetry it generates, and the induced velocity from the vortex pair, all of which amplify the deflection well beyond what a uniform shift in separation angle alone would produce. The dimple-induced near-surface destabilisation of the boundary layer is the initiating event of this chain: without it, separation would be nearly symmetric, and neither the pressure-driven sidewash nor the coherent vortex pair would develop. Figure~\ref{fig:illustration} illustrates the formation of counter-rotating vortices via the pressure-driven sidewash that eventually leads to the lift generation.

The non-monotonic dependence of lift on $k/d$ in the present simulations follows directly from this mechanism. Increasing $k/d$ from $0.004$ to $0.006$ raises $\langle C_L\rangle$, while the trend reverses at $k/d=0.008$ (Table~\ref{tab:force_comp1}). At $k/d=0.006$, the dimple-triggered transition is more effective across the dimpled hemisphere, producing a larger pressure asymmetry and stronger sidewash than at $k/d=0.004$, as reflected in the smaller separation region observed for $k/d=0.006$ (Sec.~\ref{sec3sub1}). At $k/d=0.008$, deeper dimples impose a growing roughness penalty that moves the separation point upstream, weakening the pressure asymmetry and reducing the lift \cite{Beratlis2019, vilumbrales2025}. The same competition between the transition-promoting effect of the dimples and the roughness-induced pressure drag penalty explains the experimentally observed $C_L$--$k/d$ curve across a wider range of dimple depths \cite{sudarsana2024}. The experiments also show an analogous dependence on $Re$ at fixed $k/d$: below a critical $Re$ no lift is produced, and above it $C_L$ rises sharply before saturating \cite{sudarsana2024}. The critical $Re$ decreases with increasing $k/d$ because the controlling parameter is the ratio of the dimple depth to the local boundary layer thickness. Either increasing $k/d$ at fixed $Re$ or increasing $Re$ at fixed $k/d$ can cross the threshold at which the dimples generate sufficient perturbation to trigger transition.

The wake response on the asymmetrically dimpled sphere shares one element with spinning spheres \cite{Kim2014, Milner2025} and cricket balls \cite{Parekh2024}: pressure asymmetry from differential separation drives cross-stream flow that rolls up into counter-rotating streamwise vortex pairs, which deflect the wake via their mutual induced velocity. Parekh et al.~\cite{Parekh2024} and Milner \& Scobie~\cite{Milner2025} termed these structures wing-tip-like vortices and demonstrated this amplification for seam- and rotation-induced asymmetry, respectively. The present results show that the same wake-side amplification operates on a non-rotating sphere with distributed surface roughness, indicating that the counter-rotating vortex pair is a common amplifier of wake deflection on spheres with asymmetric separation. The upstream mechanism by which the boundary-layer asymmetry develops, however, distinguishes the asymmetric-dimple case from rotation- and seam-induced asymmetry.

The key distinction of the asymmetrically dimpled sphere lies in the boundary-layer physics that produces the asymmetry. On a spinning sphere, the velocity asymmetry from rotation delays separation on the retreating side while both boundary layers remain laminar, and the lateral force arises from differential laminar separation \cite{Muto2012, Kim2014}. On a cricket ball, the seam acts as a localised trip that produces a laminar separation bubble on the seam side, with the boundary layer reattaching as turbulent and separating further downstream than on the non-seam side \cite{Parekh2024}. On the asymmetrically dimpled sphere, the distributed dimples drive a different transition pathway: the dimples destabilise the near-wall boundary layer and produce a near-surface transition over a large portion of the dimpled hemisphere \cite{Choi2006, Beratlis2019}, without an intermediate separation bubble, consistent with the near-wall TKE distribution and the post-separation concentration of $\mathcal{P}_k$ reported in Sec.~\ref{sec3sub2}. This near-surface transition produces a spatially varying three-dimensional separation topology that is absent in the localised seam effect. Two qualitatively different transition pathways therefore coexist on the same body: the dimpled side undergoes near-surface transition, while the smooth side transitions in the free shear layer downstream of laminar separation.


\section{Summary and Conclusions}\label{sec4}
The present study analyzes forces, surface flow behavior, and wake characteristics in flow past a sphere with asymmetrically dimpled surface at a Reynolds number of $Re=100{,}000$ using wall-resolved large eddy simulation. The simulations considered a smooth sphere and asymmetrically dimpled spheres with dimpled depth ratios of $k/d=0.004$, $0.006$, and $0.008$, corresponding to the regime in which substantial transverse-force generation was observed experimentally \cite{sudarsana2024}. This work highlights a new perspective into how asymmetric boundary layer perturbations over a three-dimensional bluff body modify separation topology, near-wake dynamics, and turbulence characteristics. Consistent with the experimental work, the wall-resolved LES in the present study demonstrates that a sphere with asymmetric roughness induces a finite mean lift force coefficient, $\langle C_L \rangle \approx 0.32-0.36$, while mean drag remains nearly unchanged across different $k/d$. 

The near-invariance of drag is elaborated by the pressure force decomposition. In all cases, pressure drag accounts for approximately $98\%$ of the total drag. Although the asymmetric dimples substantially redistribute the surface pressure between the front and rear hemispheres, the net streamwise-projected pressure force changes only weakly. A Fourier-based decomposition of the surface pressure further shows that pressure drag is predominantly governed by the azimuthally averaged, axisymmetric pressure component, whereas the pressure lift is governed by the non-axisymmetric pressure component. Therefore, the asymmetric dimples introduce a strong transverse pressure imbalance without producing a comparable change in the streamwise pressure projection. This provides a direct explanation for how substantial lift can be generated without a significant drag penalty. 

Asymmetric dimples on a sphere also alter the three-dimensional separation topology. For the smooth sphere, the mean separation angle remains nearly axisymmetric at $\phi_s\sim80^\circ$, consistent with subcritical smooth sphere behavior. For the asymmetrically dimpled sphere, the smooth hemisphere retains a similar laminar separation behavior, whereas the dimpled hemisphere exhibits non-uniform delayed separation, with separation angles extending to $\phi_s\sim105^\circ-125^\circ$. This spatially varying separation pattern is coupled to the pressure difference between the smooth and dimpled hemispheres, which drives a sidewash from the smooth side toward the dimpled side. The case $k/d=0.006$ which produces the largest lift, exhibits the strongest pressure asymmetry, the smallest near-wall separated region on the dimpled side, and the largest wake deflection. The wake analysis shows that lift generation is not solely as the result of a local difference in separation angle. Instead, the asymmetric pressure field drives a cross-stream sidewash that rolls up into a coherent counter-rotating streamwise vortex pair downstream of the sphere. This vortex pair induces a downward wake deflection, opposite to the direction of the lift force, and amplifies the global wake asymmetry beyond what would be expected from separation angle differences alone. The wake deflection angle follows the generated lift force trend, reaching its highest value for $k/d=0.006$. In contrast, the dominant vortex-shedding frequency remains nearly unchanged at $St\approx0.19$ for both smooth and dimpled spheres, indicating that asymmetric dimples reorganize the mean wake direction and structure without substantially altering the primary shedding frequency.

Turbulence statistics reveal that the smooth and dimpled hemispheres undergo different transition pathways. On the smooth hemisphere, the boundary layer separates in a laminar state and transitions downstream in the free shear layer, similar to the smooth sphere case. On the dimpled hemisphere, the dimples promote near-surface transition before separation, producing higher turbulent kinetic energy and its production close to the wall. This near-wall transition delays separation on the dimpled side and establishes the asymmetric pressure field that drives sidewash, the formation of a streamwise vortex pair, and wake deflection. The lift generation mechanism therefore consists of a coupled sequence: dimple-induced near-wall transition, non-uniform separation delay, transverse pressure imbalance, pressure-driven sidewash, counter-rotating streamwise vortex pair formation, and amplified wake deflection. The non-monotonic dependence of lift on dimple depth follows from the competition between transition promotion and roughness penalty. Increasing $k/d$ from $0.004$ to $0.006$ strengthens the dimple-induced transition and increases the pressure asymmetry, thereby increasing lift. Further increasing the depth to $k/d=0.008$ reduces lift due to a roughness penalty that shifts the effective separation upstream and diminishes the favorable pressure asymmetry. This behavior is consistent with the experimental trend and suggests that an optimal roughness amplitude exists for maximizing transverse force generation.

Overall, the present results establish that lift generation by asymmetric dimpled roughness on a sphere is an intrinsically three-dimensional phenomenon governed by the coupled interaction of boundary layer transition, non-uniform separation, pressure redistribution, and coherent wake dynamics. The key finding is that asymmetric surface roughness can produce large transverse forces by modifying the non-axisymmetric pressure field while leaving the axisymmetric pressure component (and therefore the mean drag) nearly unchanged. From an applied perspective, this study provides a mechanistic foundation for designing a passive or adaptive perturbation strategy to induce controllable lateral forces without incurring drag penalties. These are directly relevant to engineering applications involving bluff bodies, including underwater spherical robots, sports aerodynamics, and flow control systems.

\section*{Acknowledgments}\label{sec5}
We acknowledge the use of the Great Lakes High-Performance Computing (HPC) cluster at the University of Michigan-Ann Arbor and the Anvil HPC cluster at Purdue University. In addition, we thank the National Science Foundation (NSF) Advanced Cyberinfrastructure Coordination Ecosystem: Services \& Support (ACCESS) program for providing access to HPC resources through credit grant number \texttt{mch250036} and \texttt{mch260047}. 

\section*{Appendix}
\subsection{Separation angles estimation based on Moore-Rott-Sears (MRS) criterion}\label{appendix1}
\begin{figure}[htbp]
    \centering
    \includegraphics[width=0.8\linewidth]{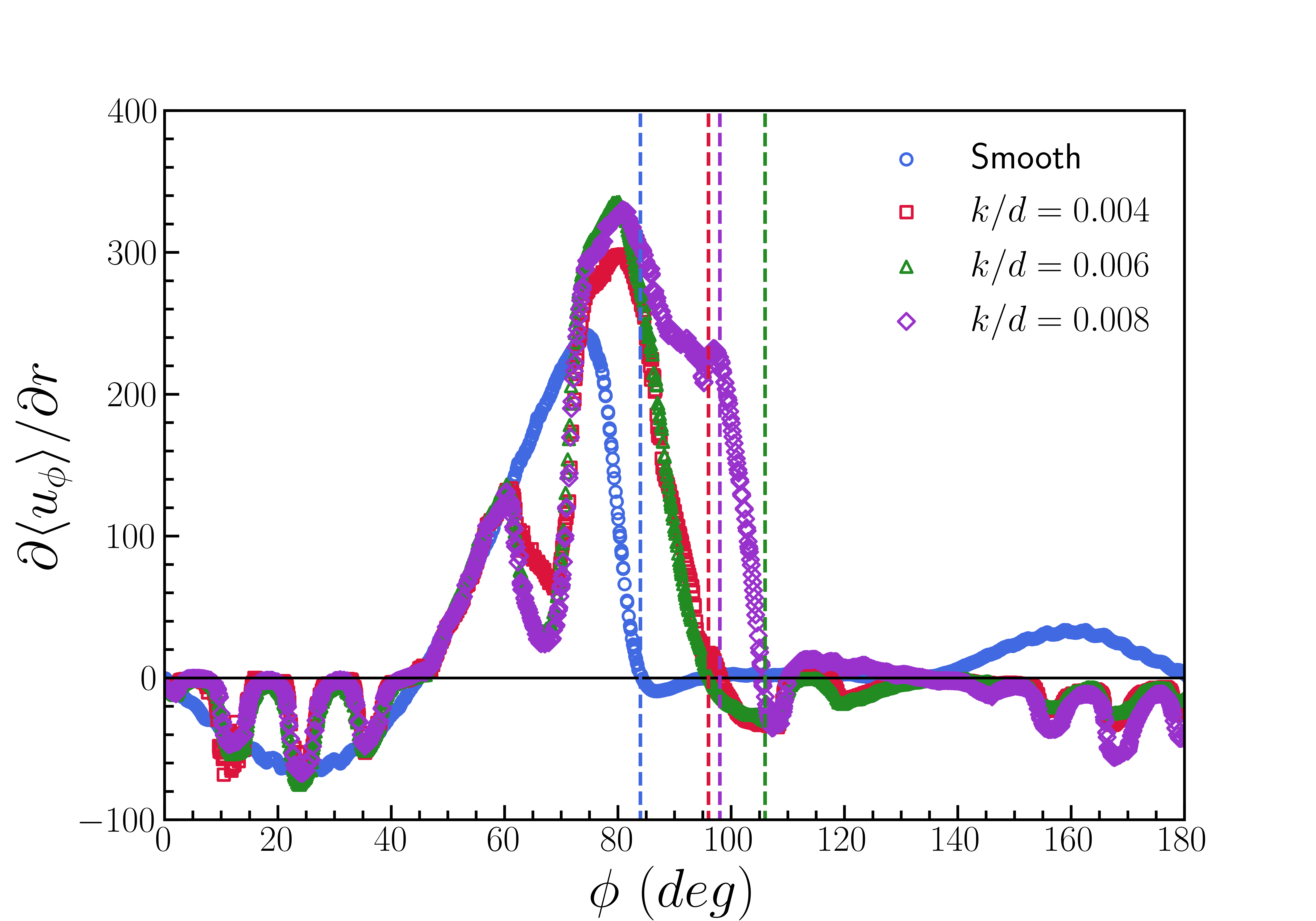}
    \caption{Global separation angle estimation via MRS Criterion $\partial \langle u_\phi \rangle /\partial r=0$ for smooth and dimpled sphere at $Re=100{,}000$. The blue, red, green, and purple dash lines show $\phi=84^\circ, 96^\circ, 106^\circ, 98^\circ $, respectively. The black horizontal solid lines denote the $\partial \langle u_\phi \rangle /\partial r=0$.}
    \label{fig:MRSgrad}
\end{figure}

\begin{figure}[htbp]
    \centering
    \includegraphics[width=0.8\linewidth]{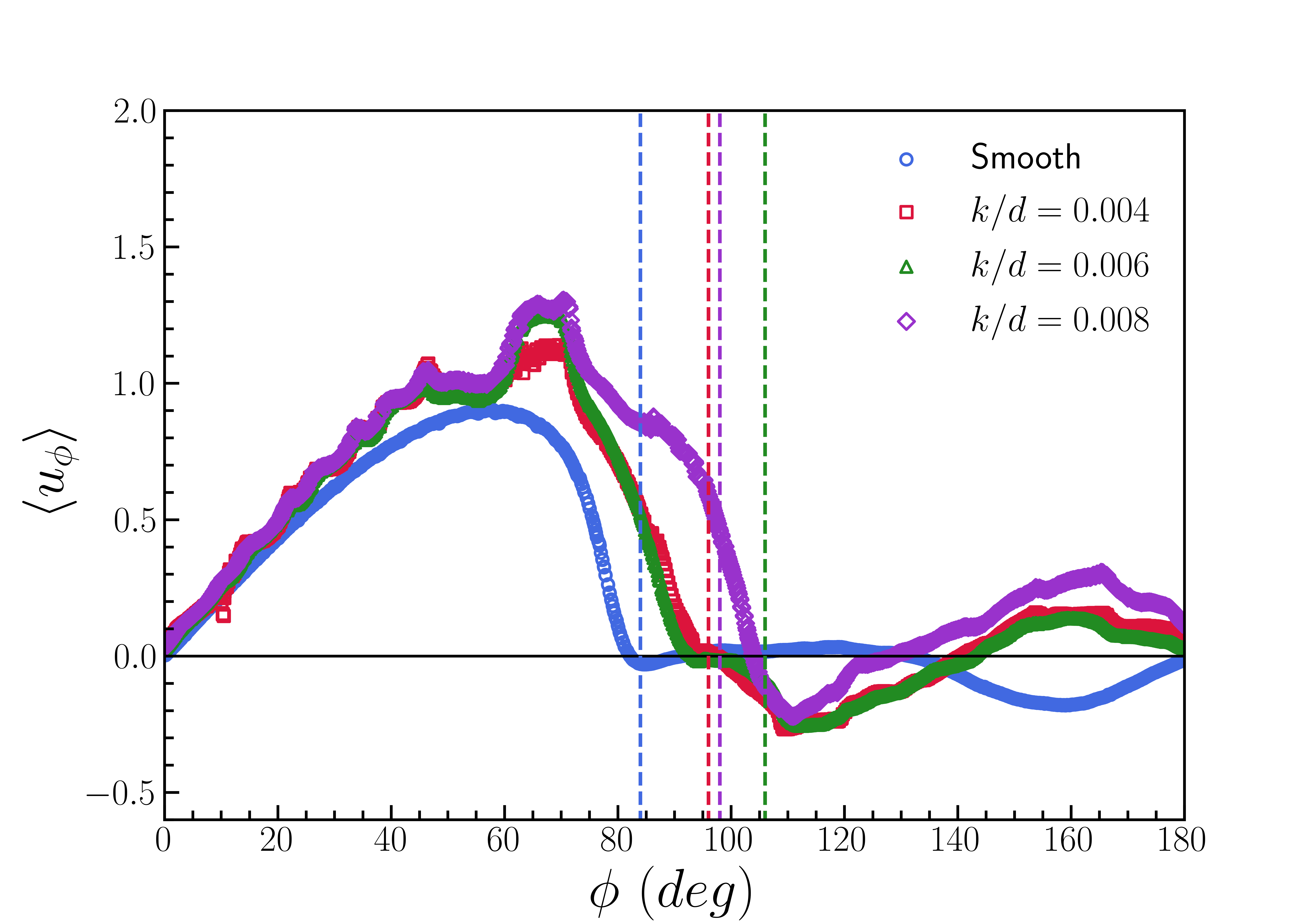}
    \caption{Global separation angle estimation via MRS Criterion $\langle u_\phi \rangle=0$ for smooth and dimpled sphere at $Re=100{,}000$. The blue, red, green, and purple dash lines show $\phi=84^\circ, 96^\circ, 106^\circ, 98^\circ $, respectively. The black horizontal solid lines denote the $\langle u_\phi \rangle\lesssim0$}
    \label{fig:MRSu}
\end{figure}

The separation angles presented in this study rely on the surface flow analysis by considering $\langle C_{f_x}\rangle \lesssim 0$ and the separation pattern criteria characterized by the topological connections of the skin-friction field on the boundary, as outlined by Lighthill et al. \cite{lighthill1963attachment} and Surana et al. \cite{Surana2006}. This is considered the classical criterion, which implies $\partial u_x/\partial y\Bigr|_{\substack{y=0}}$. This particularly allows for the estimation of separation angles across the whole surface of the sphere. Previous studies on spheres utilized the MRS criterion \cite{Kim2014, sudarsana2024, vilumbrales2025}, primarily due to its robustness as well as the practicality in an experimental setting. This criterion was developed initially for a case of steady flows over moving walls by Moore \cite{Moore1958}, Rott \cite{Rott1956UnsteadyVF}, Sears \cite{Sears1956SomeRD}, independently, and further assessed numerically by Inoue \cite{Inoue1981}. The criterion states that flow separation occurs when $\partial u_x/\partial y=0$ at $u_x=0$, or in spherical coordinates, $\partial u_\phi/\partial r=0$ at $u_\phi=0$ \cite{Kim2014}. To further validate the separation angles estimation outlined in Fig. \ref{fig:3dsepsmodim}, the classical criterion is compared with the MRS criterion at polar angle of $\theta=90^\circ$. The separation angles based on MRS criterion is obtained from the profile of $\partial \langle u_\phi \rangle /\partial r$ and $\langle u_\phi \rangle$ over azimuthal angle $\phi$ at near-wall distance of $r/d=0.502$, as shown in Figure \ref{fig:MRSgrad} and \ref{fig:MRSu}, respectively. The dash line represents the estimated angles at which $\partial \langle u_\phi \rangle /\partial r=0$ at $\langle u_\phi \rangle=0$ (Figs. \ref{fig:MRSgrad} and \ref{fig:MRSu}). For smooth and $k/d=0.004, 0.006,0.008$, the MRS criterion suggests flow separation occurs at $\phi_{MRS}\sim84^\circ, 96^\circ, 106^\circ, 98^\circ$, while the classical criterion shows at $\phi_{sep}\sim80^\circ, 103^\circ, 108^\circ, 103^\circ$. This small difference between both approaches shows that the currently employed classical criterion is valid and consistent with the MRS criterion that are widely used in flow over spheres experiments. 

\subsection{Wake deflection angle quantification}\label{appendix2}
\begin{figure}[htbp]
    \centering
    \includegraphics[width=1\linewidth]{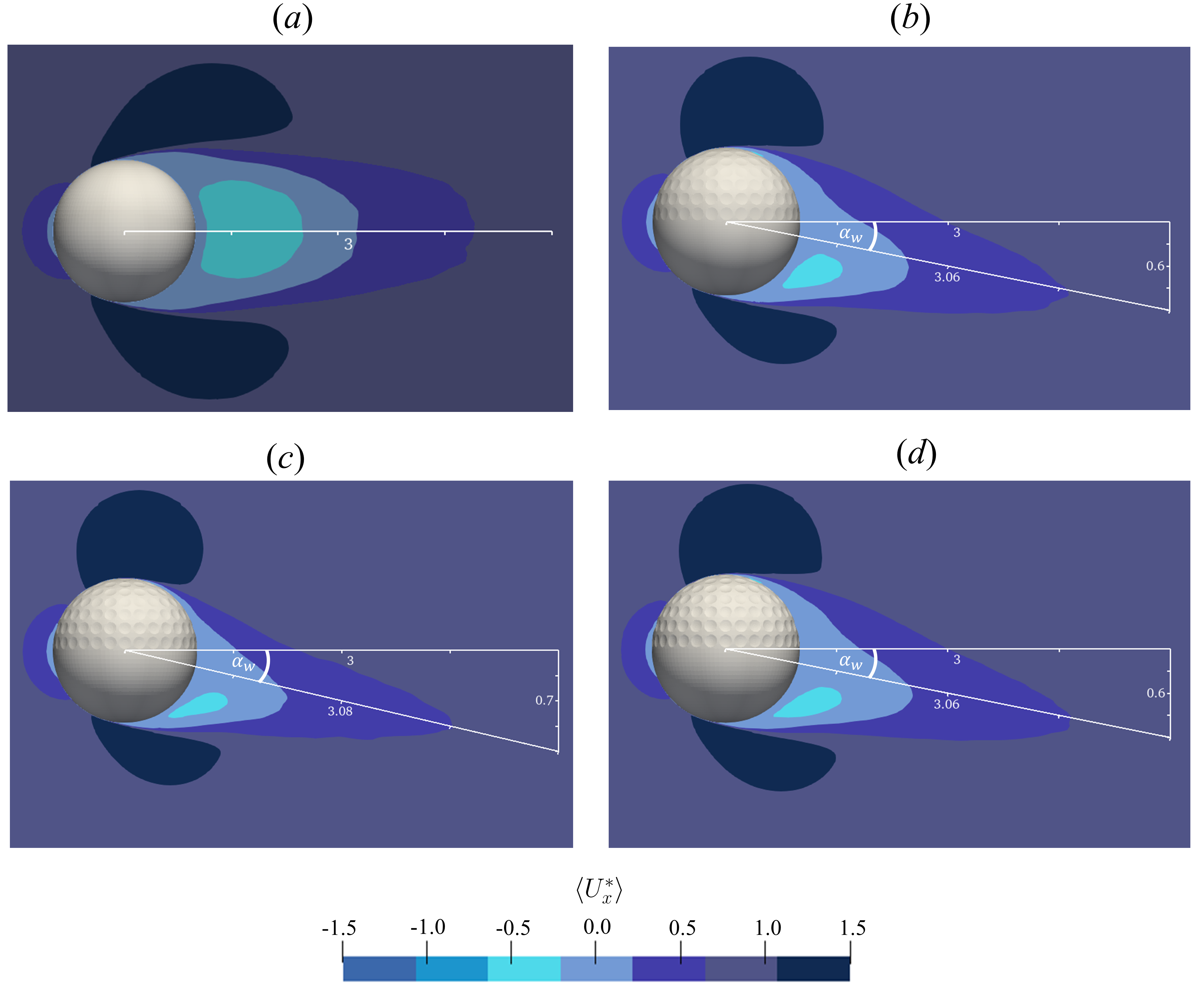}
    \caption{Wake deflection angle $\alpha_w$ estimation from the time-averaged normalized streamwise velocity $\langle U_x^*\rangle$ for (a) Smooth, (b) Asymmetric dimpled $k/d=0.004$, (c) Asymmetric dimpled $k/d=0.006$, and (d) Asymmetric dimpled $k/d=0.008$ at $Re=100{,}000$. The white lines correspond to the construction lines for estimating $\alpha_w$, annotated with its corresponding length. }
    \label{fig:wakedeflect}
\end{figure}

Quantification of the wake deflection behavior is conducted by estimating the angle $\alpha_w$ of the time-averaged streamwise velocity $\langle U_x^*\rangle$ at the midplane, as shown in Fig. \ref{fig:wakedeflect}. The white lines in the figure shows the construction lines for estimating $\alpha_w$. As expected, smooth sphere exhibits a symmetric wake structure in $+x$-axis. The wake deflection angle $\alpha_w$ for $k/d=0.004$ and $k/d=0.008$ is identical at $\sim11.3^\circ$, while $k/d=0.006$ shows the most deflected wake at $\sim13.13^\circ$, which is correlated with the $\langle C_L\rangle$ in Table \ref{tab:force_comp1}. A qualitative observation from Fig. \ref{fig:wakedeflect} suggest that the wake of $k/d=0.006$ appears thinner compared to other $k/d$ cases, also showing a consistent behavior with the $\langle C_D\rangle$ in Table \ref{tab:force_comp1}.

\newpage
\bibliographystyle{elsarticle-num} 
\bibliography{refcfdsphere}

@article{Inoue1981,
author = {Inoue, Osamu},
title = {MRS criterion for flow separation over moving walls},
journal = {AIAA Journal},
volume = {19},
number = {9},
pages = {1108-1111},
year = {1981},
doi = {10.2514/3.7848},
URL = {https://doi.org/10.2514/3.7848},
eprint = {https://doi.org/10.2514/3.7848}
}

@incollection{Moore1958,
address = {Berlin},
author = {Moore, F. K.},
booktitle = {Boundary Layer Research},
doi = {10.1016/0016-0032(59)90384-9},
isbn = {9783540022732},
issn = {00160032},
pages = {296--311},
publisher = {Springer-Verlag},
title = {{On the separation of the unsteady laminar boundary layer}},
year = {1958}
}

@article{Rott1956UnsteadyVF,
  title={Unsteady viscous flow in the vicinity of a stagnation point},
  author={Nicholas Rott},
  journal={Quarterly of Applied Mathematics},
  year={1956},
  volume={13},
  pages={444-451},
  url={https://api.semanticscholar.org/CorpusID:125371874}
}

@article{Sears1956SomeRD,
  title={Some Recent Developments in Airfoil Theory},
  author={William Rees Sears},
  journal={Journal of the Aeronautical Sciences},
  year={1956},
  volume={23},
  pages={490-499},
  url={https://api.semanticscholar.org/CorpusID:118506426}
}

@article{Achenbach1974b,
   author = {Elmar Achenbach},
   doi = {10.1017/S0022112074000644},
   issn = {14697645},
   issue = {2},
   journal = {Journal of Fluid Mechanics},
   pages = {209-221},
   title = {Vortex shedding from spheres},
   volume = {62},
   year = {1974}
}

@book{Pope2000, place={Cambridge}, title={Turbulent Flows}, publisher={Cambridge University Press}, author={Pope, Stephen B.}, year={2000}}

@article{Kim2020,
   author = {Minwoo Kim and Jiseop Lim and Seungtae Kim and Solkeun Jee and Donghun Park},
   doi = {10.1016/j.cma.2020.113287},
   issn = {00457825},
   journal = {Computer Methods in Applied Mechanics and Engineering},
   month = {11},
   publisher = {Elsevier B.V.},
   title = {Assessment of the wall-adapting local eddy-viscosity model in transitional boundary layer},
   volume = {371},
   year = {2020}
}

@article{kim2019,
title = {Large-eddy simulation with parabolized stability equations for turbulent transition using OpenFOAM},
journal = {Computers \& Fluids},
volume = {189},
pages = {108-117},
year = {2019},
issn = {0045-7930},
doi = {https://doi.org/10.1016/j.compfluid.2019.04.010},
url = {https://www.sciencedirect.com/science/article/pii/S0045793019301203},
author = {Minwoo Kim and Jiseop Lim and Seungtae Kim and Solkeun Jee and Jaeyoung Park and Donghun Park}
}

@article{qin2018,
title = {Large eddy simulations of unsteady flows over a stationary airfoil},
journal = {Computers \& Fluids},
volume = {161},
pages = {155-170},
year = {2018},
issn = {0045-7930},
doi = {https://doi.org/10.1016/j.compfluid.2017.11.014},
url = {https://www.sciencedirect.com/science/article/pii/S004579301730422X},
author = {Shiwei Qin and Manoochehr Koochesfahani and Farhad Jaberi}
}

@article{Deshpande2017,
   author = {Rahul Deshpande and Vivek Kanti and Aditya Desai and Sanjay Mittal},
   doi = {10.1017/jfm.2016.827},
   issn = {14697645},
   journal = {Journal of Fluid Mechanics},
   month = {2},
   pages = {815-840},
   publisher = {Cambridge University Press},
   title = {Intermittency of laminar separation bubble on a sphere during drag crisis},
   volume = {812},
   year = {2017}
}

@article{mehta1985,
   author = "Mehta, R. D.",
   title = "Aerodynamics of Sports Balls", 
   journal= "Annual Review of Fluid Mechanics",
   year = "1985",
   volume = "17",
   number = "Volume 17, 1985",
   pages = "151-189",
   doi = "https://doi.org/10.1146/annurev.fl.17.010185.001055",
   url = "https://www.annualreviews.org/content/journals/10.1146/annurev.fl.17.010185.001055",
   publisher = "Annual Reviews",
   issn = "1545-4479",
   type = "Journal Article",
  }

@article{Terwagne2014,
   author = {Denis Terwagne and Miha Brojan and Pedro M. Reis},
   doi = {10.1002/adma.201401403},
   issn = {15214095},
   issue = {38},
   journal = {Advanced Materials},
   month = {10},
   pages = {6608-6611},
   publisher = {Wiley-VCH Verlag},
   title = {Smart morphable surfaces for aerodynamic drag control},
   volume = {26},
   year = {2014}
}

@misc{Choi2008,
   author = {Haecheon Choi and Woo Pyung Jeon and Jinsung Kim},
   doi = {10.1146/annurev.fluid.39.050905.110149},
   isbn = {9780824307400},
   issn = {00664189},
   pages = {113-139},
   title = {Control of flow over a bluff body},
   journal = {Annual Review of Fluid Mechanics},
   volume = {40},
   year = {2008}
}

@article{Achenbach1974a,
   author = {Elmar Achenbach},
   doi = {10.1017/S0022112074001285},
   issn = {14697645},
   issue = {1},
   journal = {Journal of Fluid Mechanics},
   pages = {113-125},
   title = {The effects of surface roughness and tunnel blockage on the flow past spheres},
   volume = {65},
   year = {1974}
}

@article{Surana2006, title={Exact theory of three-dimensional flow separation. Part 1. Steady separation}, volume={564}, DOI={10.1017/S0022112006001200}, journal={Journal of Fluid Mechanics}, author={Surana, A. and Grunberg, O. and Haller, G.}, year={2006}, pages={57–103}}

@incollection{lighthill1963attachment,
  author    = {Lighthill, M. J.},
  title     = {Attachment and separation in three-dimensional flows},
  booktitle = {Laminar Boundary Layer Theory},
  editor    = {Rosenhead, L.},
  year      = {1963},
  pages     = {72--82},
  publisher = {Oxford University Press},
  chapter   = {II 2.6}
}

@article{Achenbach1972,
   author = {Elmar Achenbach},
   doi = {10.1017/S0022112072000874},
   issn = {14697645},
   issue = {3},
   journal = {Journal of Fluid Mechanics},
   month = {8},
   pages = {565-575},
   title = {Experiments on the flow past spheres at very high Reynolds numbers},
   volume = {54},
   year = {1972}
}

@article{Arya2019,
   author = {Nitish Arya and Ashoke De},
   doi = {10.1016/j.camwa.2019.03.038},
   issn = {08981221},
   issue = {6},
   journal = {Computers and Mathematics with Applications},
   month = {9},
   pages = {2035-2051},
   publisher = {Elsevier Ltd},
   title = {Effect of grid sensitivity on the performance of wall adapting SGS models for LES of swirling and separating–reattaching flows},
   volume = {78},
   year = {2019}
}

@article{Nicoud1999,
   author = {F Nicoud and F Ducros},
   journal = {Flow, Turbulence and Combustion},
   pages = {183-200},
   title = {Subgrid-Scale Stress Modelling Based on the Square of the Velocity Gradient Tensor},
   volume = {62},
   year = {1999}
}

@article{Deshpande2018,
   author = {Rahul Deshpande and Ravi Shakya and Sanjay Mittal},
   doi = {10.1017/jfm.2018.474},
   issn = {14697645},
   journal = {Journal of Fluid Mechanics},
   month = {9},
   pages = {50-82},
   publisher = {Cambridge University Press},
   title = {The role of the seam in the swing of a cricket ball},
   volume = {851},
   year = {2018}
}

@article{Milner2025,
   author = {Leo G. Milner and James A. Scobie},
   doi = {10.1017/jfm.2025.10617},
   issn = {14697645},
   journal = {Journal of Fluid Mechanics},
   month = {9},
   publisher = {Cambridge University Press},
   title = {Ordinary and inverse Magnus effects on rotating spheres: Laminar separation bubble, secondary vortex and wing-tip-like vortices},
   volume = {1019},
   year = {2025}
}

@article{Krishnan2025,
   author = {Navaneeth Krishnan and Vagesh D. Narasimhamurthy and Mahesh V. Panchagnula},
   doi = {10.1063/5.0281815},
   issn = {10897666},
   issue = {7},
   journal = {Physics of Fluids},
   month = {7},
   publisher = {American Institute of Physics},
   title = {Laminar separation bubble in a spinning ball aerodynamics},
   volume = {37},
   year = {2025}
}

@article{Muto2012,
   author = {Masaya Muto and Makoto Tsubokura and Nobuyuki Oshima},
   doi = {10.1063/1.3673571},
   issn = {10706631},
   issue = {1},
   journal = {Physics of Fluids},
   month = {1},
   publisher = {American Institute of Physics Inc.},
   title = {Negative Magnus lift on a rotating sphere at around the critical Reynolds number},
   volume = {24},
   year = {2012}
}

@article{Kim2014,
   author = {Jooha Kim and Haecheon Choi and Hyungmin Park and Jung Yul Yoo},
   doi = {10.1017/jfm.2014.428},
   issn = {14697645},
   journal = {Journal of Fluid Mechanics},
   month = {9},
   pages = {R2},
   publisher = {Cambridge University Press},
   title = {Inverse Magnus effect on a rotating sphere: When and why},
   volume = {754},
   year = {2014}
}

@article{Parekh2024,
   author = {Aman Parekh and Daksh Chaplot and Sanjay Mittal},
   doi = {10.1017/jfm.2024.135},
   issn = {14697645},
   journal = {Journal of Fluid Mechanics},
   month = {3},
   publisher = {Cambridge University Press},
   title = {Swing and reverse swing of a cricket ball: Laminar separation bubble, secondary vortex and wing-tip-like vortices},
   volume = {983},
   year = {2024}
}

@article{Beratlis2019,
   author = {Nikolaos Beratlis and Elias Balaras and Kyle Squires},
   doi = {10.1017/jfm.2019.647},
   issn = {14697645},
   journal = {Journal of Fluid Mechanics},
   month = {11},
   pages = {147-167},
   publisher = {Cambridge University Press},
   title = {On the origin of the drag force on dimpled spheres},
   volume = {879},
   year = {2019}
}

@article{Choi2006,
   author = {Jin Choi and Woo Pyung Jeon and Haecheon Choi},
   doi = {10.1063/1.2191848},
   issn = {10706631},
   issue = {4},
   journal = {Physics of Fluids},
   publisher = {American Institute of Physics Inc.},
   title = {Mechanism of drag reduction by dimples on a sphere},
   volume = {18},
   year = {2006}
}

@article{Chae2026,
   author = {Seokbong Chae and Yujin Kim and Jooha Kim},
   doi = {10.1017/jfm.2025.11075},
   issn = {14697645},
   journal = {Journal of Fluid Mechanics},
   month = {1},
   publisher = {Cambridge University Press},
   title = {Passive drag reduction on a sphere using azimuthally spaced surface protrusions: effects of protrusion number at fixed coverage},
   volume = {1026},
   year = {2026}
}

@article{Beratlis2012,
   author = {Nikolaos Beratlis and Kyle Squires and Elias Balaras},
   doi = {10.1080/14685248.2012.676182},
   issn = {14685248},
   journal = {Journal of Turbulence},
   pages = {1-15},
   publisher = {Taylor and Francis Ltd.},
   title = {Numerical investigation of Magnus effect on dimpled spheres},
   volume = {13},
   year = {2012}
}

@article{Issa1986,
title = {Solution of the implicitly discretised fluid flow equations by operator-splitting},
journal = {Journal of Computational Physics},
volume = {62},
number = {1},
pages = {40-65},
year = {1986},
issn = {0021-9991},
doi = {https://doi.org/10.1016/0021-9991(86)90099-9},
url = {https://www.sciencedirect.com/science/article/pii/0021999186900999},
author = {R.I Issa}
}

@article{Patankar1972,
title = {A calculation procedure for heat, mass and momentum transfer in three-dimensional parabolic flows},
journal = {International Journal of Heat and Mass Transfer},
volume = {15},
number = {10},
pages = {1787-1806},
year = {1972},
issn = {0017-9310},
doi = {https://doi.org/10.1016/0017-9310(72)90054-3},
url = {https://www.sciencedirect.com/science/article/pii/0017931072900543},
author = {S.V Patankar and D.B Spalding}
}

@article{vilumbrales2025,
  title={Adaptive drag reduction of a sphere using smart morphable skin},
  author={Vilumbrales-Garcia, Rodrigo and Sudarsana, Putu Brahmanda and Sareen, Anchal},
  journal={Flow},
  volume={5},
  pages={E17},
  year={2025},
  publisher={Cambridge University Press},
  doi={10.1017/flo.2025.7}
}

@article{sudarsana2024,
  title={On the lift generation over a sphere using asymmetric roughness},
  author={Sudarsana, Putu Brahmanda and Vilumbrales-Garcia, Rodrigo and Sareen, Anchal},
  journal={Physics of Fluids},
  volume={36},
  number={12},
  year={2024},
  publisher={AIP Publishing},
  doi={10.1063/5.0241948}
}

@article{Jeong_Hussain_1995, title={On the identification of a vortex}, volume={285}, DOI={10.1017/S0022112095000462}, journal={Journal of Fluid Mechanics}, author={Jeong, Jinhee and Hussain, Fazle}, year={1995}, pages={69–94}}

@inproceedings{Hunt1988,
  author    = {Hunt, J. C. R. and Wray, A. A. and Moin, P.},
  title     = {Eddies, streams, and convergence zones in turbulent flows},
  booktitle = {Studying Turbulence Using Numerical Simulation Databases, 2. Proceedings of the 1988 Summer Program},
  year      = {1988},
  month     = {December},
  publisher = {Legacy CDMS},
  note      = {Work of the US Gov. Public Use Permitted.},
  url       = {https://ntrs.nasa.gov/citations/19890015184} 
}

@misc{openfoam2406,
  author       = {{OpenFOAM Foundation}},
  title        = {{OpenFOAM v2406 Released}},
  howpublished = {\url{https://www.openfoam.com/news/main-news/openfoam-v2406}},
  year         = {2024},
  note         = {Accessed: 2025-04-15}
}

\end{document}